\documentclass[useAMS,usenatbib]{mn2e}

\usepackage[dvips]{graphicx}

\title[Multiple tidal tails around satellite systems]
{Development of multiple tidal tails around globular clusters
and dwarf satellite galaxies}

\author[S. Hozumi and A. Burkert]
{Shunsuke Hozumi$^1$\thanks{E-mail: hozumi@edu.shiga-u.ac.jp (SH);
burkert@usm.lmu.de (AB)}
 and Andreas Burkert$^{2,3,4\,\star}$\\
$^1$Faculty of Education, Shiga University, 2-5-1 Hiratsu,
Otsu, Shiga 520-0862, Japan\\
$^2$Universit\"atssternwarte der Ludwig-Maximilians Universit\"at,
Scheinerstr.\ 1, D-81679 M\"unchen, Germany\\
$^3$Max-Planck-Institut f\"ur Extraterrestrische Physik,
Giessenbachstr.\ 1, D-85748 Garching, Germany\\
$^4$Max-Planck-Fellow}

\begin{document}

\date{Accepted 2014 October 29;
      Received 2014 September 7; in original form 2014 August 11}

\pagerange{\pageref{firstpage}--\pageref{lastpage}}
\pubyear{2014}

\maketitle

\label{firstpage}

\begin{abstract}
The formation and evolution of tidal tails like those observed
around some globular clusters and dwarf satellite galaxies is
examined with an $N$-body simulation. In particular, we analyse
in detail the evolving tidal features of a one-component satellite
that is moving on a highly eccentric orbit in the external field
of a host galaxy potential like our own. The results show that
every time the satellite approaches apogalacticon, a fresh pair
of tidal tails becomes notably prominent, and that eventually,
the satellite possesses multiple tidal tails via repeating
apocentre passages. Accordingly, the number of observed tidal
arms can be used as a tracer of the number of orbital periods
that such a system has completed around the centre of its host
galaxy. By identifying the arm particles included in each of
the first three consecutively formed pairs of tidal tails,
we find that each pair of tidal tails is practically identical
to one another regarding the energy and angular-momentum
distributions. In addition, we demonstrate that the density
profiles of these three pairs of tidal tails at their first
apogalacticons after formation agree well with one another.
It therefore follows that the multiplicity in tidal features
originates from the repeated episode of tidal-arm formation
in the course of the precessing motion of a satellite.
\end{abstract}


\section{Introduction}
Recently, the observational evidence for tidal tails around
globular clusters (GCs) has been growing since \citet*{grillmair95}
pointed out wing-like structures in the surface-density profiles
of GCs as a signature of tidal tails. Subsequent observations
have confirmed and reinforced \citet{grillmair95} findings
\citep*[amongst many others,][]{lehmann97,
testa00, lmc00, siegel01, lee03, sollima11, balbinot11}, providing
further pieces of evidence that support the existence of tidal
tails associated with GCs. Above all, Palomar 5 is well-known for
its two long and massive tidal tails that extend to the opposite
directions from the main body \citep{oden01, oden02, oden03},
and NGC 5466 is also known to have extended giant tails
\citep{belokurov06, gj06}. These two GCs are representative
examples of ongoing tidally disrupting stellar systems in the
Milky Way.

Indications of tidal features have also been observed in
some dwarf galaxies. For instance, among Milky Way satellites,
a tidal extension was found in Ursa Minor \citep{martinez01a},
and an excess distribution of stars induced by tidal interactions
was detected around Carina \citep{munoz06} and Leo I \citep{sohn07}.
As another example, the surface brightness around Fornax was shown
to depend on the direction of measurement \citep{coleman05}, which
is reminiscent of tidal distortion. Even outside the Milky Way,
strong tidal features have been discovered e.g. around the
Andromeda satellite, NGC 205 \citep{howley08,sav10} or around
a dwarf galaxy in the Hydra I galaxy cluster that belongs to
the Local Volume \citep{rich12, koch12}. Furthermore, a dwarf
galaxy in the act of tidal disruption has been reported in the
halo of NGC 4449 which is also a dwarf galaxy in the nearby
universe \citep{martinez12}.

As described above, tidal tails are common features of satellite
systems moving around a host galaxy. In such a situation, the
force field of the host in which a satellite is orbiting will
lead to the formation of a variety of tidal tails around the
satellite. Consequently, we can estimate the distribution of
the gravitational potential of the host galaxy from the morphology
of the tidal tails \citep[e.g.][]{binney08}. Since Galactic GCs
are, on average, distributed much closer to the centre of our
Galaxy than satellite dwarf galaxies, the tidal tails of GCs
can give an indication of the inner mass profile of our Galaxy,
while those of satellite dwarf galaxies can be used as a probe
for the outer mass distribution. In addition, if the morphology
of tidal tails could help to trace the orbit of a GC or a dwarf
galaxy, we might be able to obtain an assembly history of
satellite systems through which a galaxy is built up and
eventually becomes a large and massive grown-up. Thus,
through the tidal tails of satellite systems, we could verify
observationally the hierarchical structure formation scenario
\citep[e.g.][]{bk10}.

From the viewpoints mentioned above, many numerical simulations
have been carried out so far regarding the tidal interactions
between a satellite system and its host galaxy \citep*{piatek95,
grillmair98, mateoetal98, martinez01a, martinez01b, martinez04,
dehnen04, capuzzo05, montuori07, cwk07, pen09, klkmmp09, kuep12,
lgk13}. Some studies have focused on the connection between the
orientation of tidal tails and the satellite orbit in order to
examine whether the orbital path of a satellite can be determined
from the observed morphology of its tidal tails \citep{capuzzo05,
montuori07, klkmmp09, lgk13}. Unfortunately, the orientation
of the tails is not a direct indicator of the orbital path,
because it depends on the orbital phase: the tidal tails start
to develop with the orientation almost parallel to the orbital
path near the pericentre and the orientation changes gradually
over time to become perpendicular near the apocentre if the orbit
is relatively eccentric. However, as stated by \citet{montuori07},
the tails located quite far from the GC centre are good tracers
of the GC path. This is because the tidal tails continue to
extend along both sides of the cluster path with time, once
the cluster has passed the first apocentre after the tails
were extracted near the pericentre.

Although considerable attention has been paid to the
direction of tidal tails around satellite systems, there are
no sufficient descriptions of the complicated features of tidal
tails that are forming while a satellite system is moving on
an eccentric orbit. In fact, intricate features like multiple
tidal arms can be seen in the simulations of \citet{dehnen04},
\citet{capuzzo05}, and \citet{montuori07}. \citet{montuori07}
noted the peculiar morphology of multiple arms when the cluster
was approaching the apocentre. They explained the difference in
the direction of tidal arms at the pericentre and apocentre
based on the direction and strength of the Coriolis acceleration
at both positions. Indeed, their explanation describes how the
direction of a given pair of tidal arms changes as its main
body moves on an eccentric orbit. However, how the multiplicity
is formed during the orbital motion has never been elucidated
exactly.

In this paper, we aim at disclosing the formation mechanism
of multiple tidal tails around a satellite using an $N$-body
simulation. In so doing, we pay attention only to collisionless
systems and we will ignore the internal evolution caused by
two-body relaxation in the case of collisional systems like GCs.
However, as the two-body relaxation time-scale is typically long
compared to an orbital period, we expect collisional effects in
GCs to be negligible, especially in the outer regions where the
tidal tails develop. In order to keep the analyses simple, we
also neglect the fact that dwarf spheroidal galaxies are in
effect two-component systems, consisting of a stellar inner
spheroid embedded in a surrounding dark-halo component. This
is justified by the fact that dwarf spheroidals are characterised
by large mass-to-light ratios \citep{mateo98, gilmore07},
indicating that the self-gravity of the stellar component is
negligible. Dwarf spheroidals therefore to a good approximation
can be treated as a one-component system with the stars tracing
the inner part of the tidally distorted dark-matter component.

In Section 2, we describe our model of a satellite and its
host galaxy, and the $N$-body simulation method, together with
the initial setup of the simulation. For convenience, we will
call our satellite a dwarf galaxy and we will adopt typical
physical parameters that apply to such systems. Note, however,
that all parameters can be rescaled to correspond to systems
with smaller masses so as to be adjusted more suitably for
GCs. Results are presented in Section 3. In Section 4, we
discuss the formation mechanism of the multiple tidal tails
and the detectability of such tidal tails. Finally, conclusions
are given in Section 5.

\section{Models and Method}
We follow the dynamical evolution of a satellite
that is tidally interacting with its host galaxy.
In particular, our objective is to understand the mechanism
by which multiple tidal tails develop in the course of the
orbital motion of the satellite. For this purpose, we choose
a single typical setup that will cause the
multiplicity in tidal tails, and analyse its formation process
in detail.

\begin{figure*}
\centerline{\includegraphics[width=10cm]{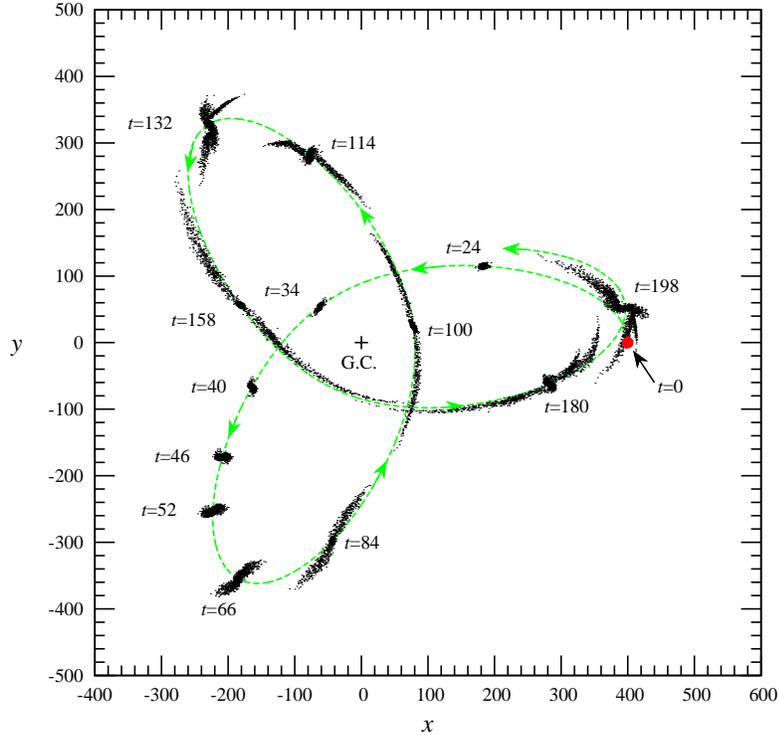}}
\caption{Time evolution of the dwarf satellite galaxy in
the act of tidal interaction with its host galaxy from the
beginning till $t=198$. The particle distribution is seen
in the orbital $(x-y)$ plane. At times denoted by the labels,
one-tenth of all particles at each position of the dwarf
galaxy are randomly selected and displayed. The filled red
circle at $(400, 0)$ stands for the initial position of the
dwarf galaxy, and the $+$ symbol represents the centre of
the host galaxy. The dashed green line shows the orbital
path of the dwarf galaxy that moves in the directions of
the arrows. The dwarf galaxy is represented
on real scales.}
\label{fig:evolution}
\end{figure*}

The host galaxy is represented as an external potential that
consists of an infinitely extended isothermal sphere. Its
potential, $\Phi$, is written as
\begin{equation}
\Phi(r)={v_\rmn{c}}^2\ln(r/r_0),
\label{eq:host_pot}
\end{equation}
where $r$ is the radial distance from the centre of the host
galaxy, $r_0$ is a reference radius which, for conventional
reasons, ensures negative values of the potential over a practical
range of radii, and $v_\rmn{c}$ is a constant circular velocity.
This host galaxy corresponds to an extremely simplified model
of a dark-matter halo that leads to the flat rotation curve
often observed in disc galaxies \citep{sofue01}.
In this model,
a disc component is not included, so that we do not take into
account the effects of disc shocking that would be suffered
by a satellite when it penetrates the disc, and those of
likely precession of the satellite orbit that would arise
from the resulting asymmetric gravitational field. Since
the pericentre distance of our dwarf galaxy model is 20
kpc as described below, these effects could be negligibly
small, which means that we could be allowed to neglect
them. However, if our adopted parameters are rescaled
to be adjusted to GCs, a disc component would be needed
to evaluate these effects. Although the simulation is
scale free, for better demonstration we adopt $r_0=125$
kpc and $v_\rmn{c}=220\ \rmn{km\,s}^{-1}$ to mimic a
typical real Milky Way-type galaxy.

The satellite is modelled as a Plummer sphere
\citep{plummer11} whose density, $\rho_\rmn{d}$, and potential,
$\Phi_\rmn{d}$, are, respectively, given by
\begin{equation}
\rho_\rmn{d}(r')=\frac{3M}{4\pi a^3}[1+(r'/a)^2]^{-5/2},
\end{equation}
and
\begin{equation}
\Phi_\rmn{d}(r')=-\frac{GM}{a}[1+(r'/a)^2]^{-1/2},
\end{equation}
where $G$ is the gravitational constant, $M$
and $a$ are the mass and scale length of the satellite,
respectively, and $r'$ stands for the radial distance from
its centre. We realise this model with $N=50\,000$ particles
of equal mass by truncating it at $r'=5\,a$. As a result, its
total mass amounts to 94.3 per cent of the corresponding
perfect Plummer sphere. For the satellite constructed here,
however, $M$ is assigned to this finite model. We use $M=10^7$
M$_\odot$ and $a=0.25$ kpc as typical values, corresponding to a
dwarf satellite galaxy. This satellite galaxy is put into an orbit
with an eccentricity of two-thirds, the value of which is suitable
for the formation of multiple tidal arms \citep{dehnen04, montuori07},
as well as typical for most dwarf galaxies formed in cosmological
simulations \citep{dkm07}. Here, the eccentricity, $\epsilon$, is
defined as
\begin{equation}
\epsilon = \frac{r_\rmn{a}-r_\rmn{p}}{r_\rmn{a}+r_\rmn{p}},
\end{equation}
where $r_\rmn{a}$ and $r_\rmn{p}$ are the apo- and peri-galactic
distances, respectively. We set $r_\rmn{a}$ to be 100 kpc and
$r_\rmn{p}$ to be 20 kpc, and commence a simulation by placing
the dwarf galaxy at apogalacticon. From the beginning, the full
host potential was imposed on the dwarf galaxy. This setup
procedure would not harm the entire dynamics of the dwarf
galaxy, since the crossing time, $t_\rmn{cr}$, of the dwarf
galaxy, as estimated at the half-mass radius, $r'{_\rmn{h}}$,
is approximately one-twentieth of the orbital time-scale, when
$t_{\rm cr}$ is calculated as $t_\rmn{cr}=r'{_\rmn{h}}/\sigma_0$,
where $\sigma_0$ is the one-component velocity dispersion at
$r'{_\rmn{h}}$.

The simulation was carried out with a hierarchical tree
algorithm \citep{bh86}. Forces were computed with an
opening angle of 0.5 and a gravitational softening of
$0.068\,a$, including terms up to quadrupole order in
the multipole expansions. The equations of motion were
integrated in Cartesian coordinates with a time-centred
leap-frog algorithm \citep[e.g.][]{press86}. Results
are presented in the system of units such that $G=M=a=1$.
Thus, when scaled to physical values, unit time and velocity
are $1.86\times 10^7$ yr and 13.1 km$\,\rmn{s}^{-1}$,
respectively. The simulation was continued until time
$t=500$ with a constant time step of 0.05, which guaranteed
that the total energy of the system was conserved to better
than $1.67\times 10^{-4}$.

\section{Results}
\begin{figure}
\centerline{\includegraphics[width=7cm]{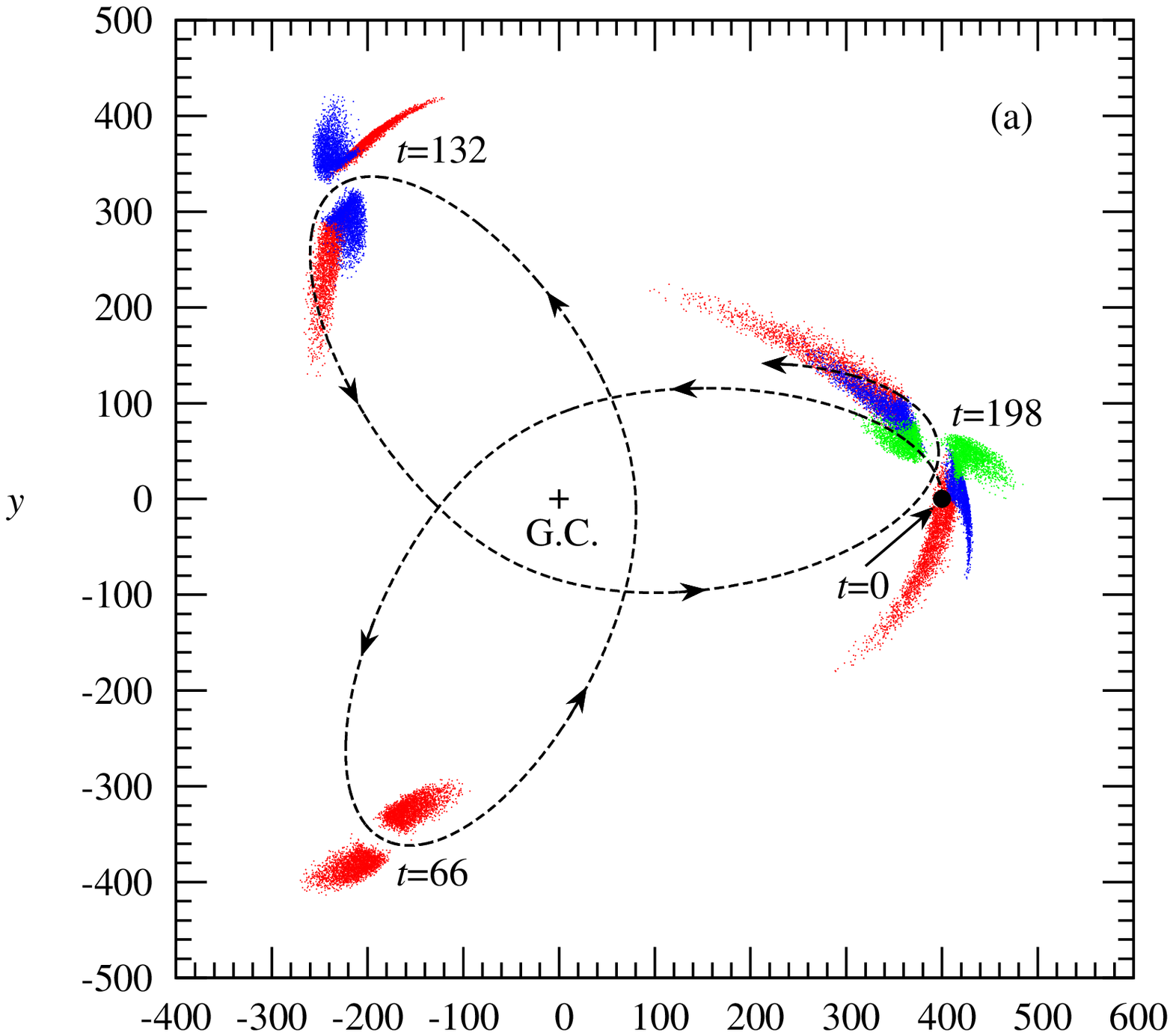}}
\centerline{\includegraphics[width=7cm]{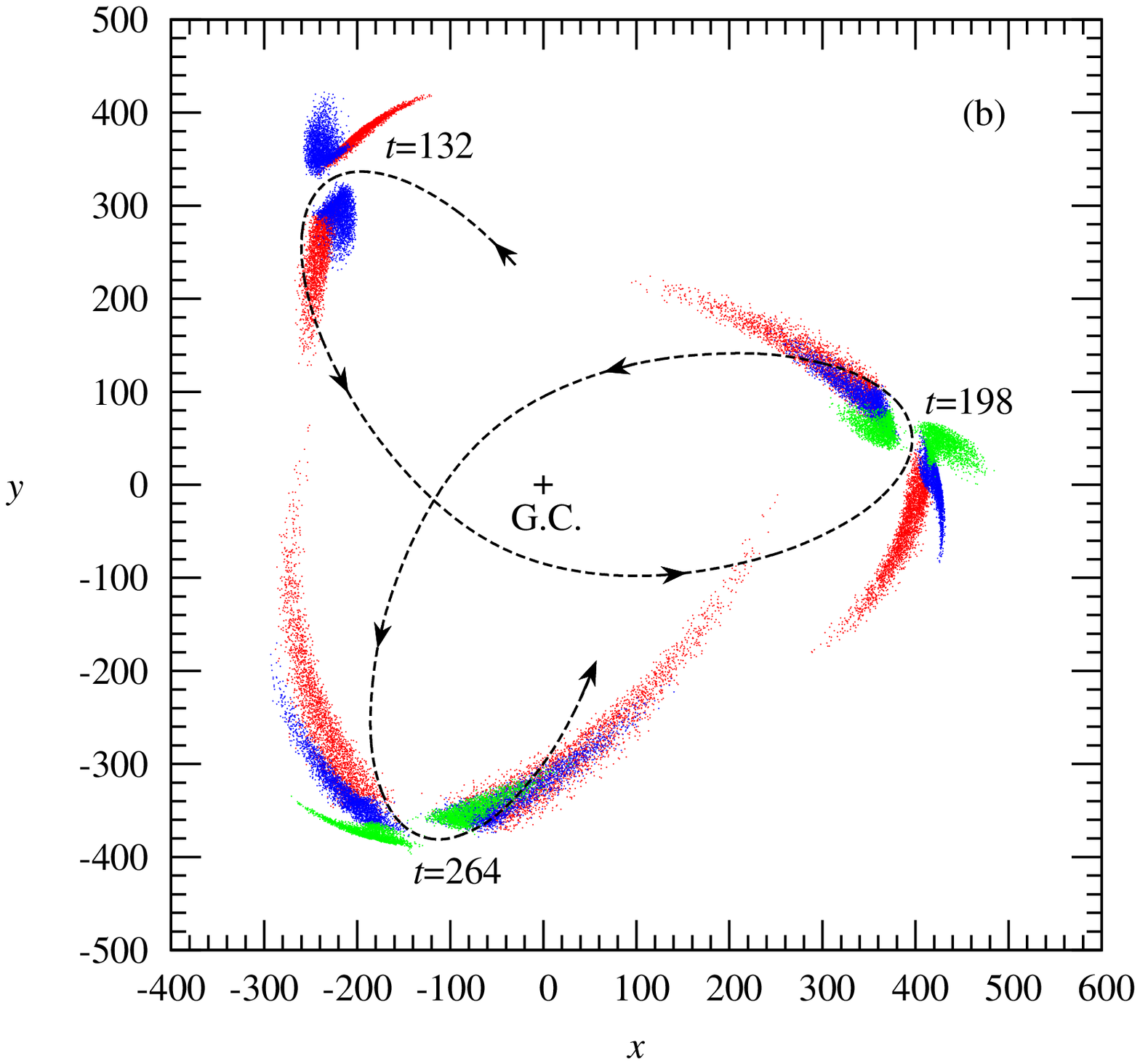}}
\caption{Forming and evolving phases of tidal tails seen in
the orbital plane from $t=0$ to $t=198$ (a), and from $t=132$
to $t=264$ (b). Only the particles included in each of the
first three consecutively formed pairs of tidal arms are
shown without depicting the main body of the dwarf galaxy.
In reality, the fourth pair of tidal arms is formed at $t=264$,
which is omitted in this plot. The red, blue, and green dots
represent the particles of the tidal arms formed firstly,
secondly, and thirdly, respectively, from the beginning of
the simulation. The dashed lines trace the orbit of the
centre of the dwarf galaxy, with the arrows indicating the
moving directions of the orbit. The filled circle at $(400, 0)$
in (a) denotes the initial position of the dwarf galaxy.
The $+$ symbols indicate the centre of the host galaxy.
Note that the tidal tails are magnified
twice as large as in reality for convenience in illustration.}
\label{fig:tails}
\end{figure}

\begin{figure}
\centerline{\includegraphics[width=7cm]{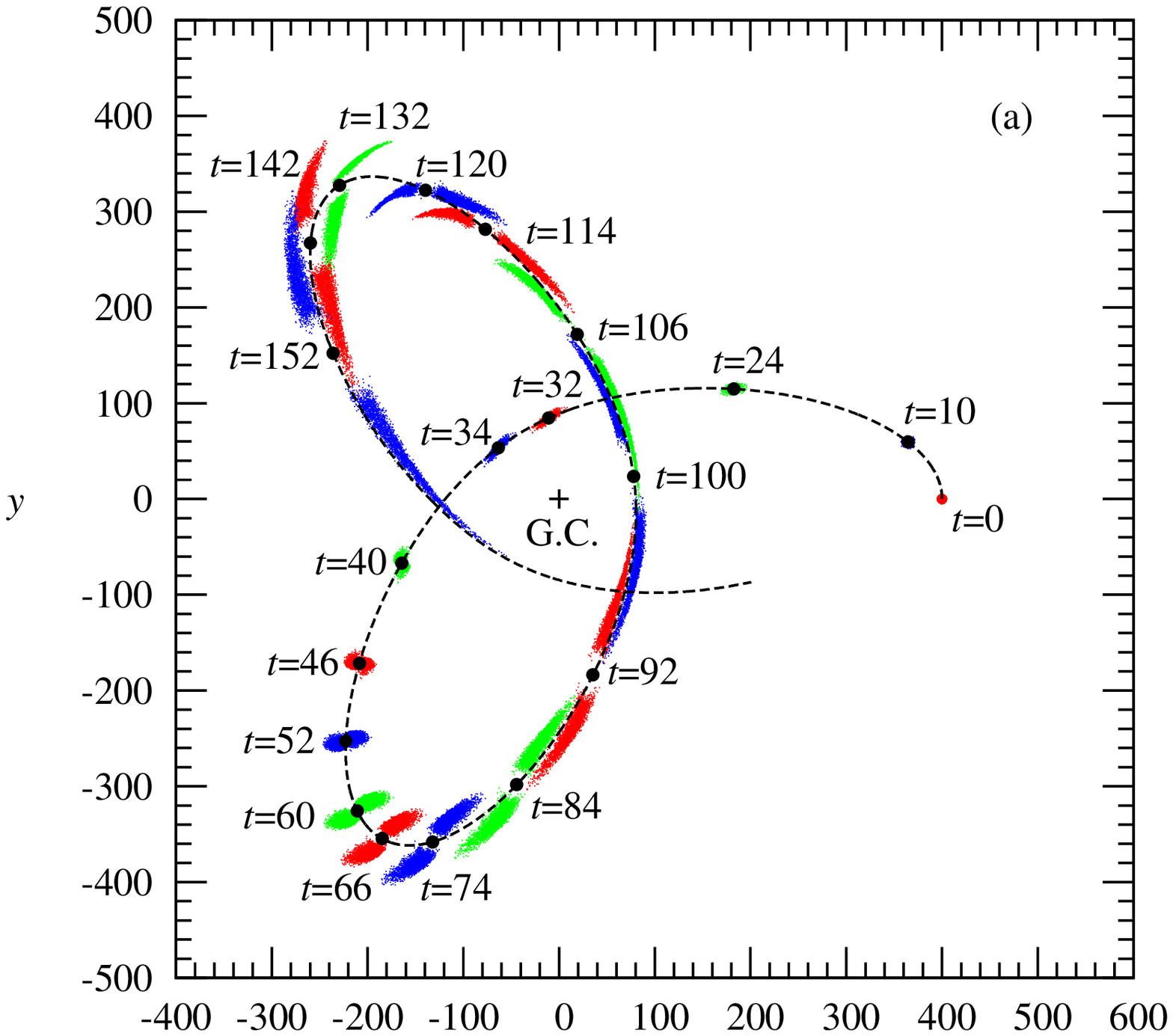}}
\centerline{\includegraphics[width=7cm]{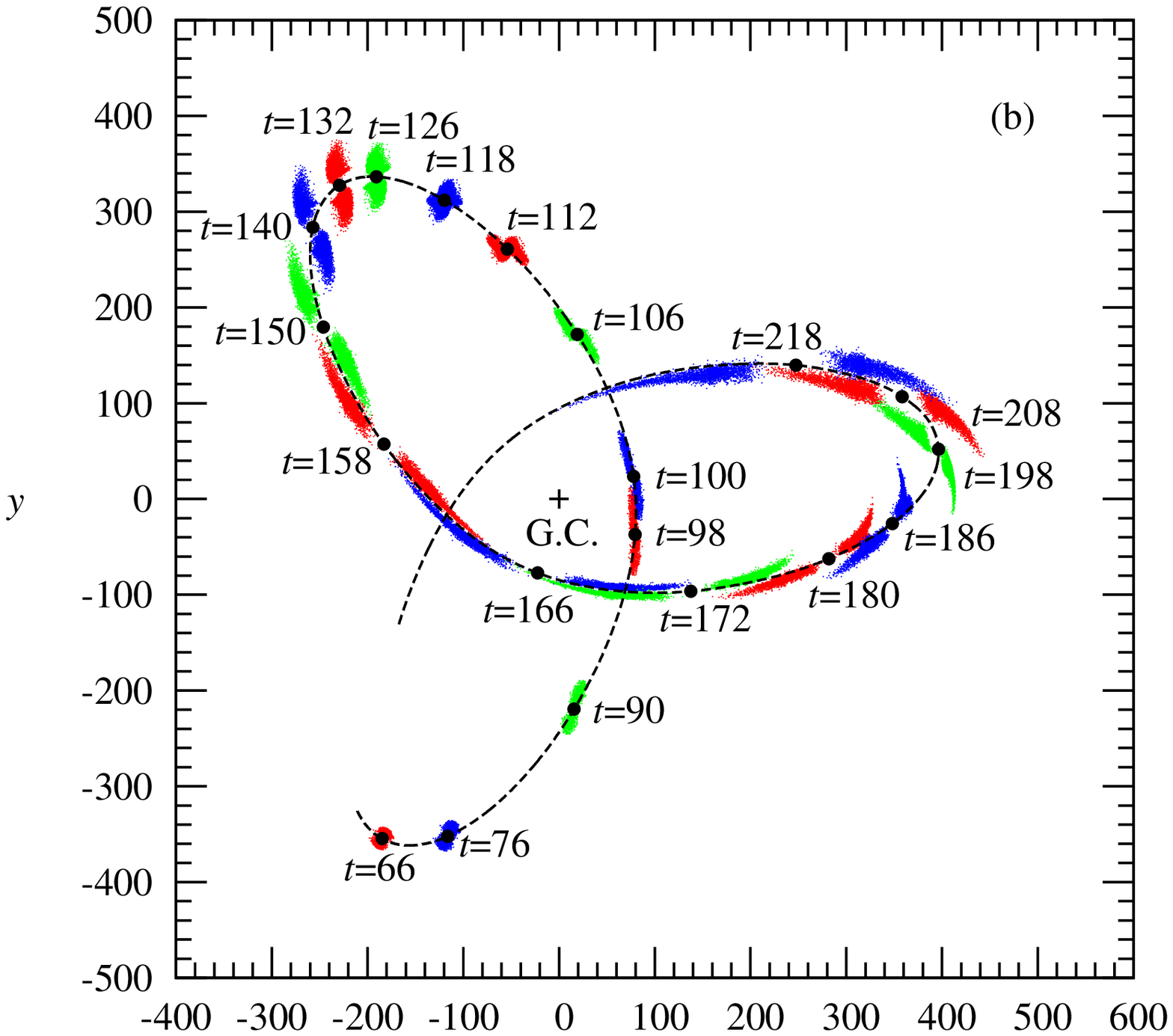}}
\centerline{\includegraphics[width=7cm]{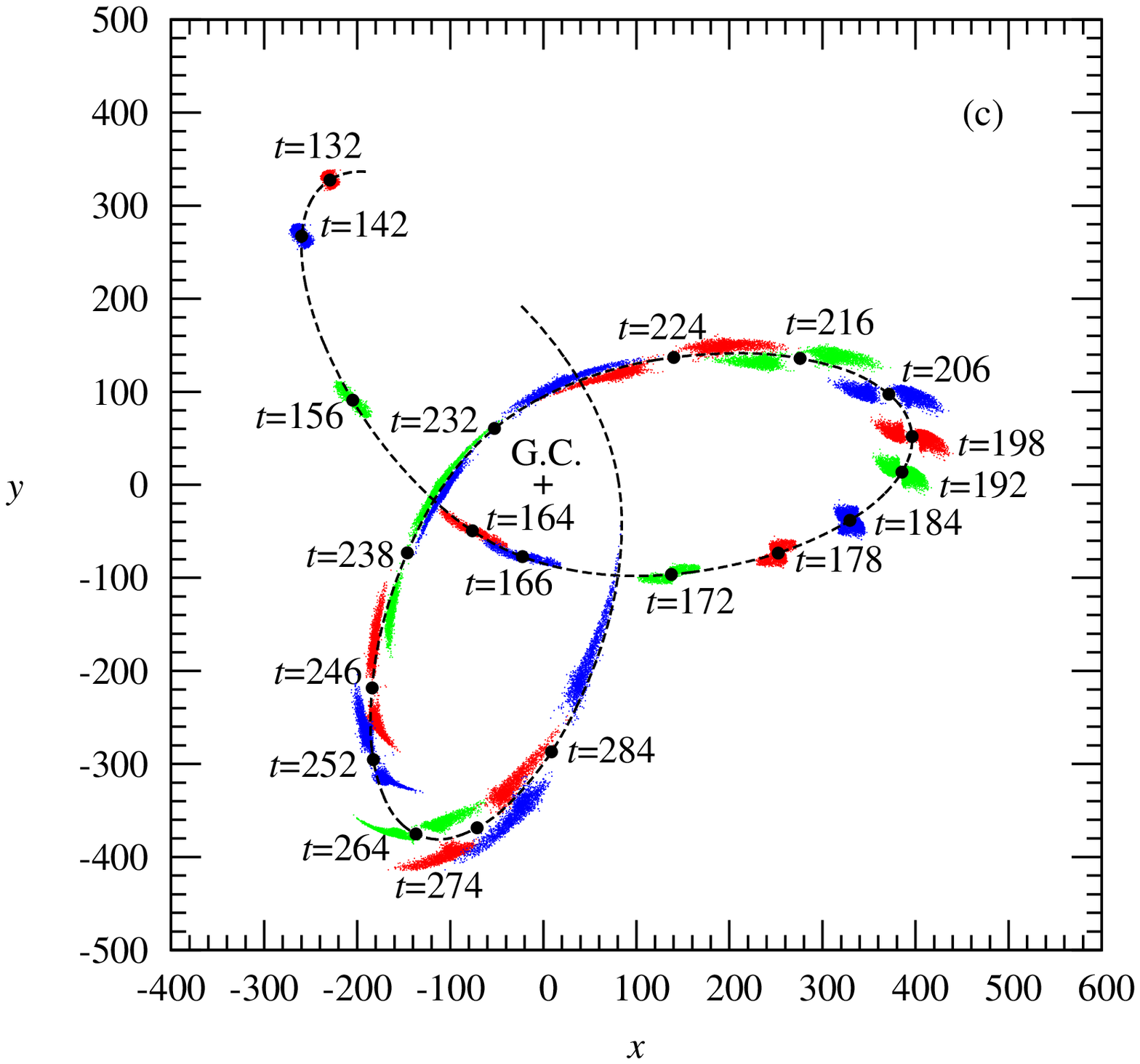}}
\caption{Time evolution of the tidal arms formed firstly
(a), secondly (b), and thirdly (c) seen in the orbital
plane. Again, only the particles included in each pair
of tidal arms are displayed. The filled circles denote
the centre of the dwarf galaxy at times shown by the
labels. In each panel, the starting time for plotting
arms is very close to apogalacticon, and the subsequent
plotting time intervals are the same for the three pairs
of tidal arms. The three colours of red, blue, and green
are repeatedly used in this order with time just for the
purpose of facilitating visualisation. The tidal tails
are depicted on real scales unlike those in
Fig.~\ref{fig:tails}.}
\label{fig:armevolution}
\end{figure}

Fig.~\ref{fig:evolution} shows the time evolution of the
dwarf galaxy that is suffering tidal forces from its host
galaxy. This figure reveals the formation process of tidal
tails, with the dwarf galaxy being heavily deformed over
time. In particular, we can see that tidal tails became
progressively very intricate. Since the principal aim of our
study is to unravel how these complicated tidal structures
illustrated in Fig.~\ref{fig:evolution} are formed, it is
useful to decompose the tidal arms in chronological order.
In order to distinguish the particles contained in the firstly
formed tidal tails from the main body, we select particles
which are both marginally bound to and unbound
from the main body at the first apogalacticon (at $t\sim 66$),
and then, from these particles, we determine
the particles that are unbound from the main body afterwards
all the way to the end of the simulation ($t=500$). Eventually,
we have identified the firstly formed tidal tails. Then,
repeating this procedure at the subsequent two apogalacticons
(at $t\sim 132$ and at $t\sim 198$) while eliminating particles
already confirmed as arm particles, we can single out the
particles belonging to the secondly and thirdly formed tidal
tails. Fig.~\ref{fig:tails} displays the forming and evolving
phases of the three groups of tidal arms thus classified by
virtue of the formation epoch, with the main body of the dwarf
galaxy being masked. As found from this figure, the arm particles
formed at each epoch can be separated into either leading-arm or
trailing-arm particles on the basis of their positions relative
to the main body when seen in the orbital plane.

\begin{figure}
\centerline{\includegraphics[width=8cm]{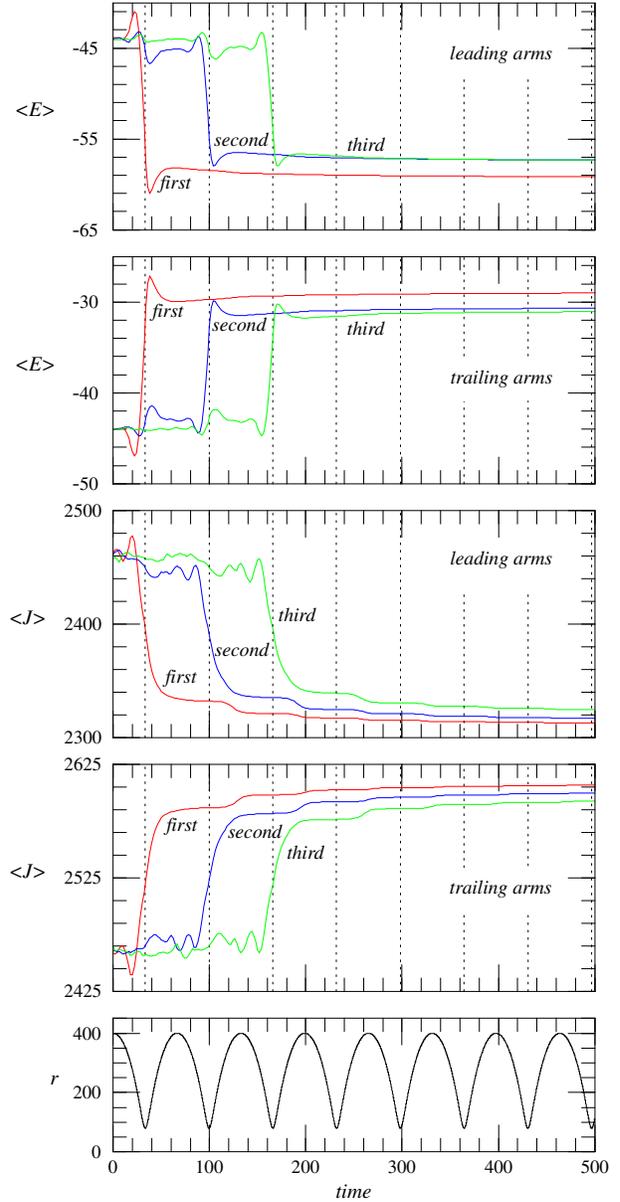}}
\caption{Time evolution of the mean specific energy,
$\langle E \rangle$, and that of the mean specific angular
momentum, $\langle J\rangle$, averaged over all particles
included in each arm. The red, blue, and green lines
represent the values of $\langle E\rangle$ or those of
$\langle J\rangle$ for the firstly, secondly, and thirdly
formed arms, respectively. The top and third panels
show the quantities for the leading arms, while the second
and fourth panels show those for the trailing arms. The
vertical dotted lines denote the times of pericentre passage.
The bottom panel indicates the radial distance between the
centres of the dwarf and host galaxies as a function of time.}
\label{fig:EandJ}
\end{figure}

\begin{figure}
\centerline{\includegraphics[width=7cm]{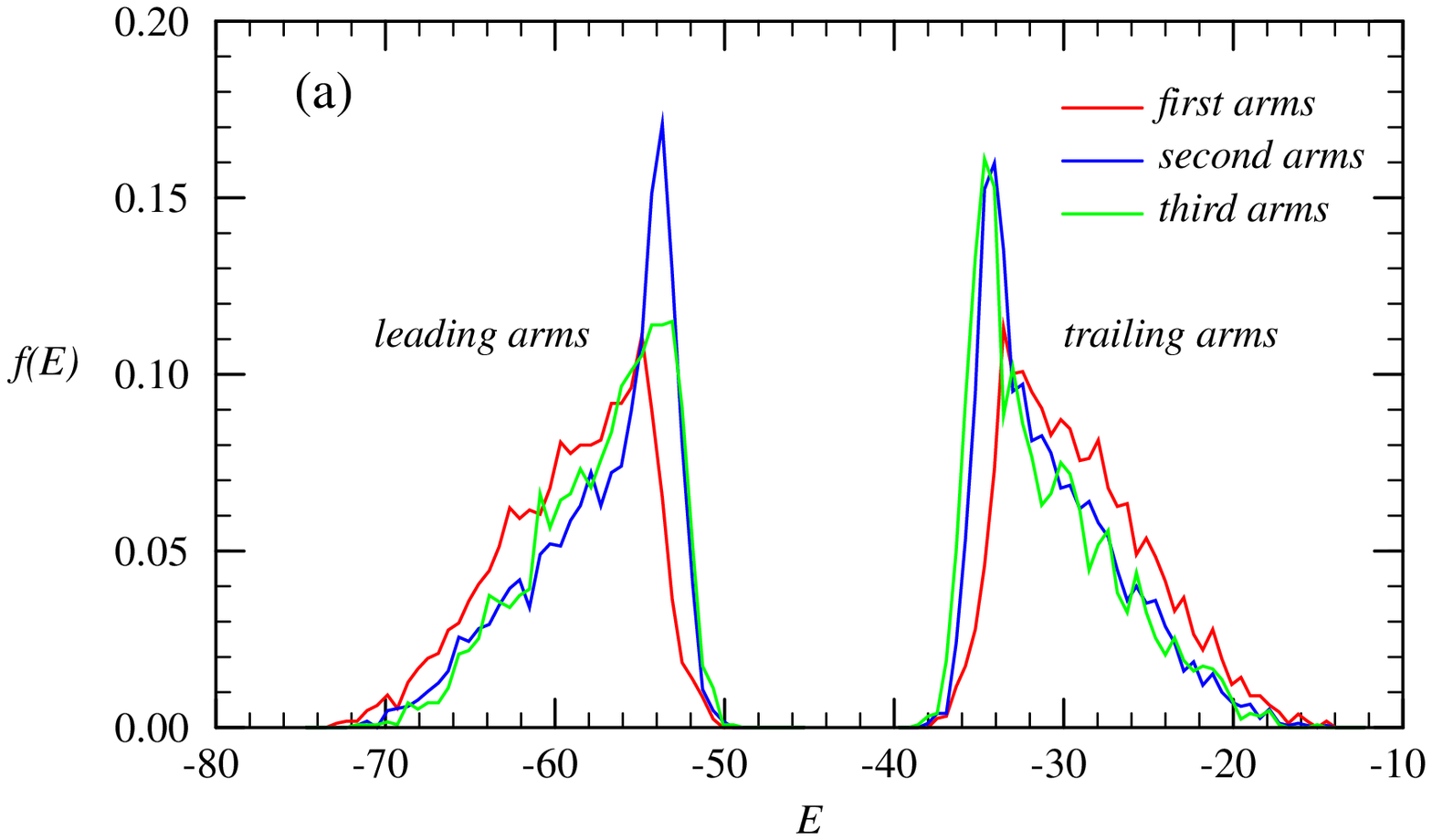}}
\vspace{5mm}
\centerline{\includegraphics[width=7.3cm]{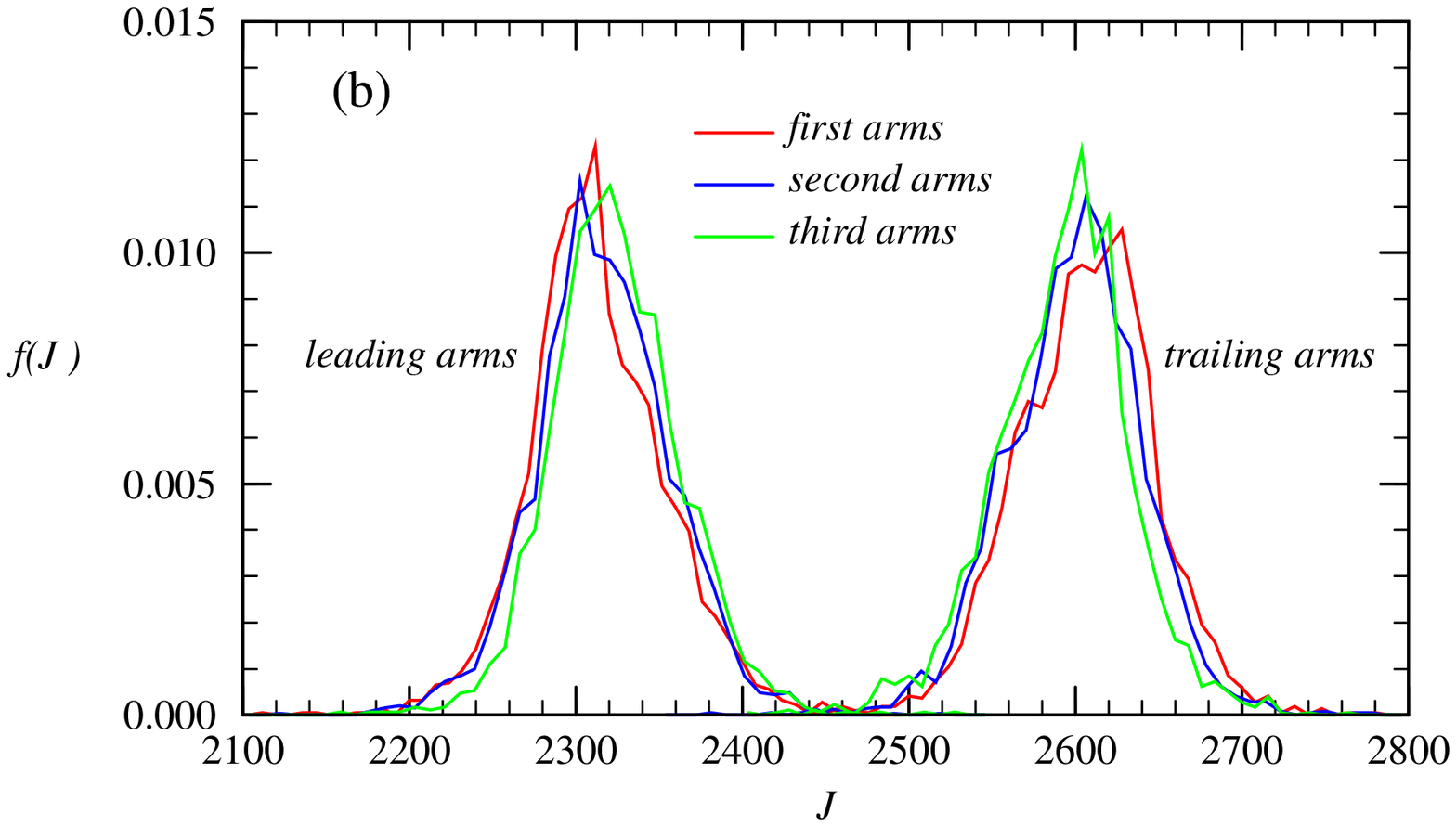}}
\caption{Fractional energy distribution, $f(E)$, (a), and
fractional angular-momentum distribution, $f(J)$, (b),
for each pair of the first three consecutively formed
tidal arms at $t=500$. The value of the energy and that
of the angular momentum at the centre of the dwarf galaxy
are $-42.2$ and $2\,466$, respectively. These distributions
remain practically unchanged after $t\sim 300$.}
\label{fig:EJ_distri}
\end{figure}

\begin{figure}
\centerline{\includegraphics[width=8cm]{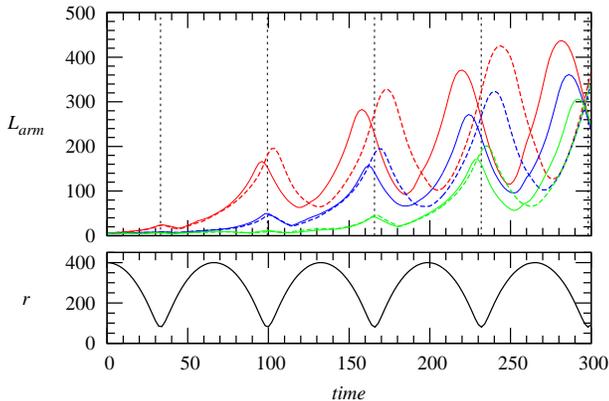}}
\caption{Time evolution of the lengths of leading (solid lines)
and trailing (dashed lines) arms (upper panel), and time change
in the radial distance between the centres of the dwarf and host
galaxies (lower panel). The length of an arm is the distance
between the nearest and farthest particles in the arm from the
centre of the dwarf galaxy. The red, blue, and green lines
represent the lengths of the firstly, secondly, and thirdly
formed tidal arms, respectively. The vertical dotted lines
indicate the times of pericentre passage.}
\label{fig:armlength}
\end{figure}

We find from Fig.~\ref{fig:tails} that a pair of tidal arms
is newly generated every time the dwarf galaxy completes a
sequential move from one apogalacticon to another, and that
the evolution of each pair of tidal arms looks very similar
to one another. For example, the pair of tidal arms at $t=66$
represented with red dots resembles the two pairs of those
which appear at $t=132$ with blue dots and at $t=198$ with
green dots. Likewise, the pair of the secondly formed tidal
tails at $t=198$ marked with blue dots has similar appearance
to the pair of those at $t=264$ with green dots.

Fig.~\ref{fig:armevolution} exhibits the time evolution of
the first three pairs of tidal arms separately, again without
depicting the main body. This figure clearly demonstrates
that the formation and evolution of the secondly formed pair
of tidal arms is one episode behind those of the firstly
formed pair, and that this relation between the firstly and
secondly formed pairs is basically repeated by the secondly
and thirdly formed pairs concerning the formation and evolution
of tidal arms. However, there are also some minor differences.
The particles contained in the first pair are distributed rather
compactly from the beginning to $t=46$ somewhat after the first
pericentre passage as compared to those in the second and third
pairs which show expanded distributions from the time around
apocentre passage ($t=66$ for the second pair and $t=132$ for
the third pair) to the time somewhat after pericentre passage
($t=112$ for the second pair and $t=178$ for the third pair).
This is because the particles in the second and third pairs
have already suffered tidal effects once (the third pair)
or twice (the second pair) in passing pericentre (see also
Fig.~\ref{fig:EandJ} below). Nevertheless, except at the
early formation phases, the appearance of the three pairs
of tidal arms at each corresponding phase looks very similar
to one another.

Fig.~\ref{fig:EandJ} shows the time evolution of the mean
specific energy, $\langle E\rangle$, and that of the mean
specific angular momentum, $\langle J\rangle$, each of which
is calculated by averaging over all particles included in
each leading or trailing arm of the first three consecutively
formed tidal tails, along with the radial distance between
the centres of the dwarf and host galaxies against time.
To be precise, the specific energy is the total energy, and
includes both potentials of the dwarf and host galaxies, while
the specific angular momentum is its absolute value being
calculated with respect to the centre of the host galaxy.
From Fig.~\ref{fig:EandJ}, we find that the dwarf galaxy
suffers strong tidal forces within the potential of the
host galaxy when it passes pericentre, which leads to the
excitation of a pair of tidal tails, and that as a consequence,
particles in trailing arms gain both energy and angular momentum,
while those in leading arms lose both energy and angular momentum.
Although all arm particles gain some energy and subsequently
become unbound from the dwarf galaxy, the total energies of
those leading-arm particles, which are moving slightly inside
the orbital path of the dwarf galaxy after being tidally
extracted, decrease by an amount which is much larger than
the gained energy owing to further energy loss that originates
from the difference in the host galaxy potential between the
particle positions before and after the tidal extraction. We
provide estimations of the energy and angular-momentum changes
for the arm particles at the formation epoch in Appendix A. In
addition, we notice from Fig.~\ref{fig:EandJ} that leading-arm
(trailing-arm) particles periodically lose (gain) angular
momentum by a tiny amount after they escape from the dwarf
galaxy and form tidal tails; a closer look at the late time
evolution in the figure indicates that the subsequent changes
in angular momentum of arm particles occur not near perigalacticon
but near apogalacticon. This angular-momentum change is caused by
the gravitational force of the main satellite body that attracts
the leading-arm particles backward and the trailing-arm particles
forward, and that the effect of the attractive force is largest
at apogalacticon where the arm particles are closest to the
satellite (see Fig.~\ref{fig:armlength}).

\begin{figure*}
\centerline{\includegraphics[width=12cm]{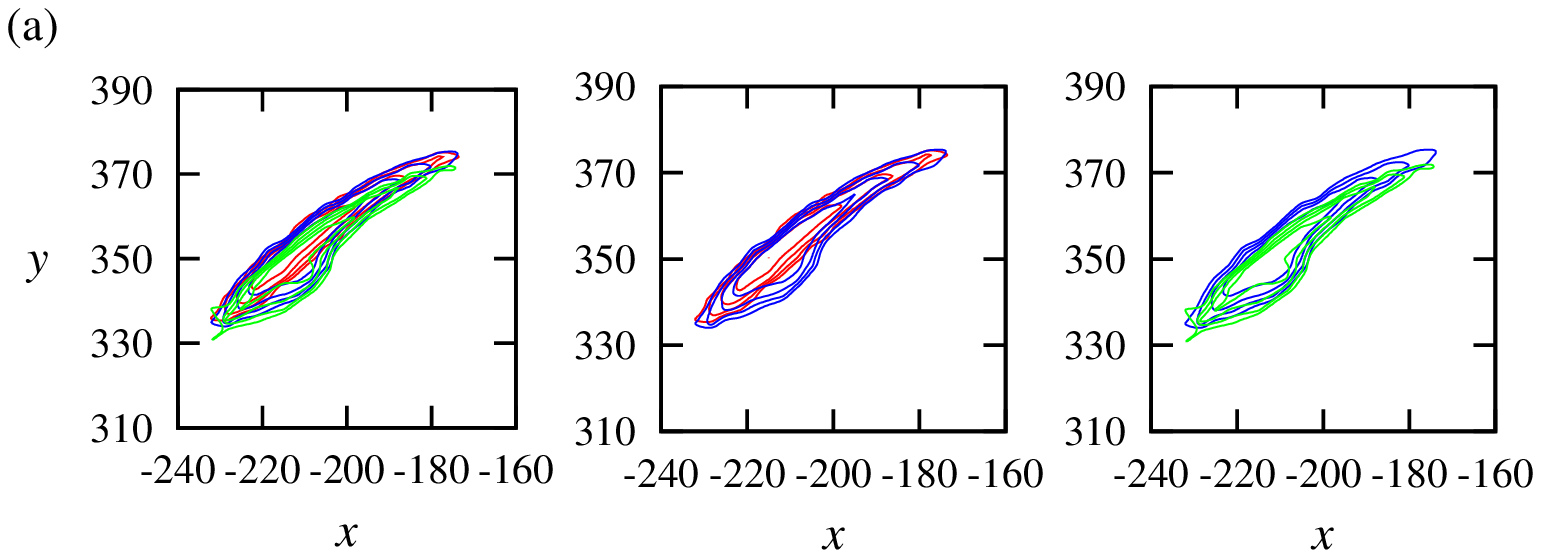}}
\includegraphics[width=12cm]{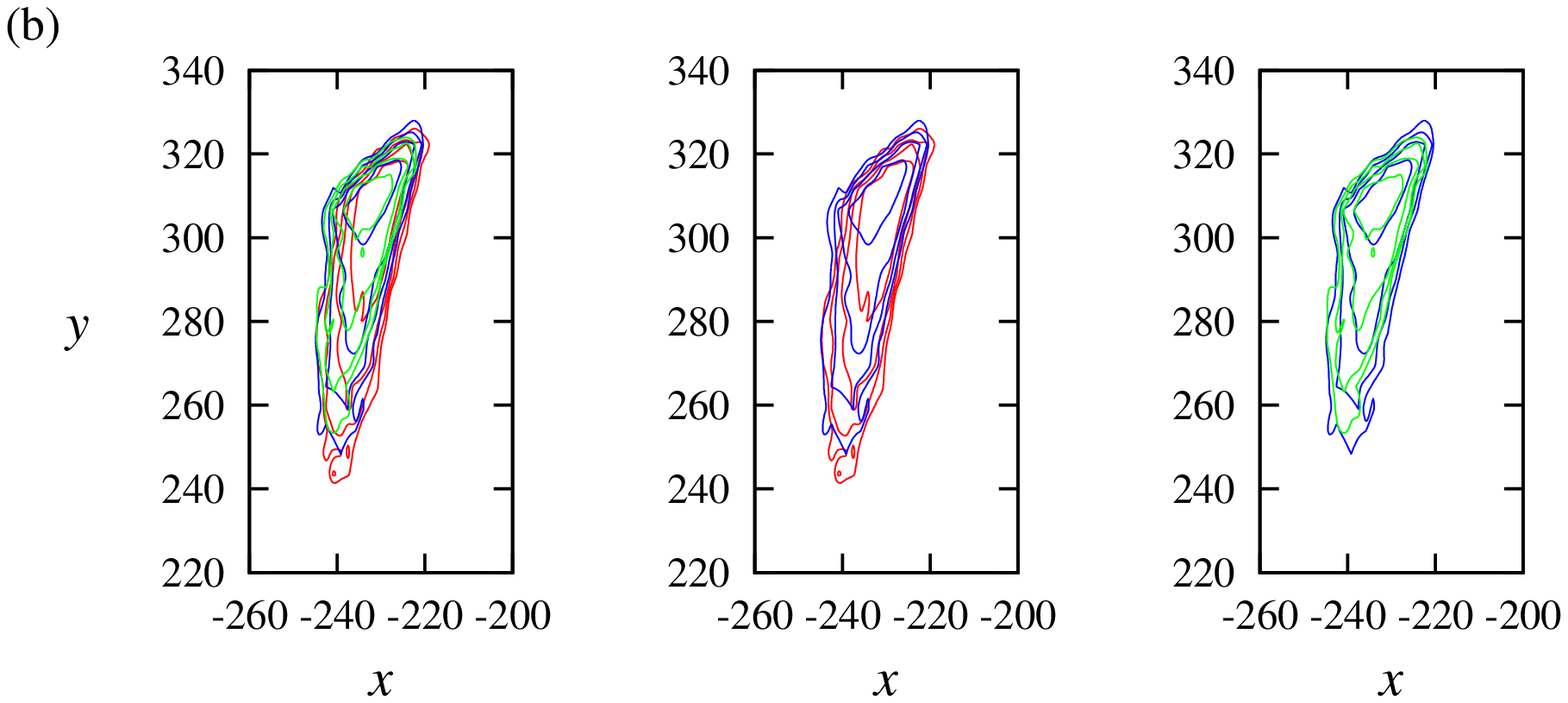}
\caption{Superposition of the density contours projected on to
the orbital plane of the first three consecutively formed tidal
tails by rotating an angle between two specified apogalacticons
for trailing arms (a) and leading arms (b). All three arms
(left panels), the first (at $t=132$) and second (at
$t=198$) arms (middle panels), and the second (at $t=198$)
and third (at $t=264$) arms (right panels) are superposed
at the positions of the first trailing and leading arms at
$t=132$. The density contours are drawn at the 80\%, 60\%,
40\%, and 20\% levels of the maximum density of each arm on
logarithmic scales. The red, blue, and green contours denote
the density contours of the first, second, and third arms,
respectively.}
\label{fig:superpose_arms}
\end{figure*}

Figs.~\ref{fig:EJ_distri}(a) and \ref{fig:EJ_distri}(b)
present, respectively, the fractional energy distribution,
$f(E)$, and the fractional angular-momentum distribution,
$f(J)$, for each pair of the first three consecutively
formed tidal arms at the end of the simulation ($t=500$).
As found from Fig.~\ref{fig:EandJ}, the arm particles
experience only a small amount of change in energy and
angular momentum after they have escaped from the potential
of the satellite, and so, the distributions of energy and
angular momentum for the first three pairs of tidal tails
can be regarded as approximately invariable if they are
examined at times somewhat later than the formation of
the third pair of tidal arms. Thus, Figs.~\ref{fig:EJ_distri}(a)
and \ref{fig:EJ_distri}(b) indicate that the energy and
angular-momentum distributions of the three pairs of tidal
arms are almost identical to one another. Particularly
remarkable is the agreement in $f(J)$ for both leading
and trailing arms among the three pairs of tidal arms.

\begin{figure}
\centerline{\includegraphics[width=7cm]{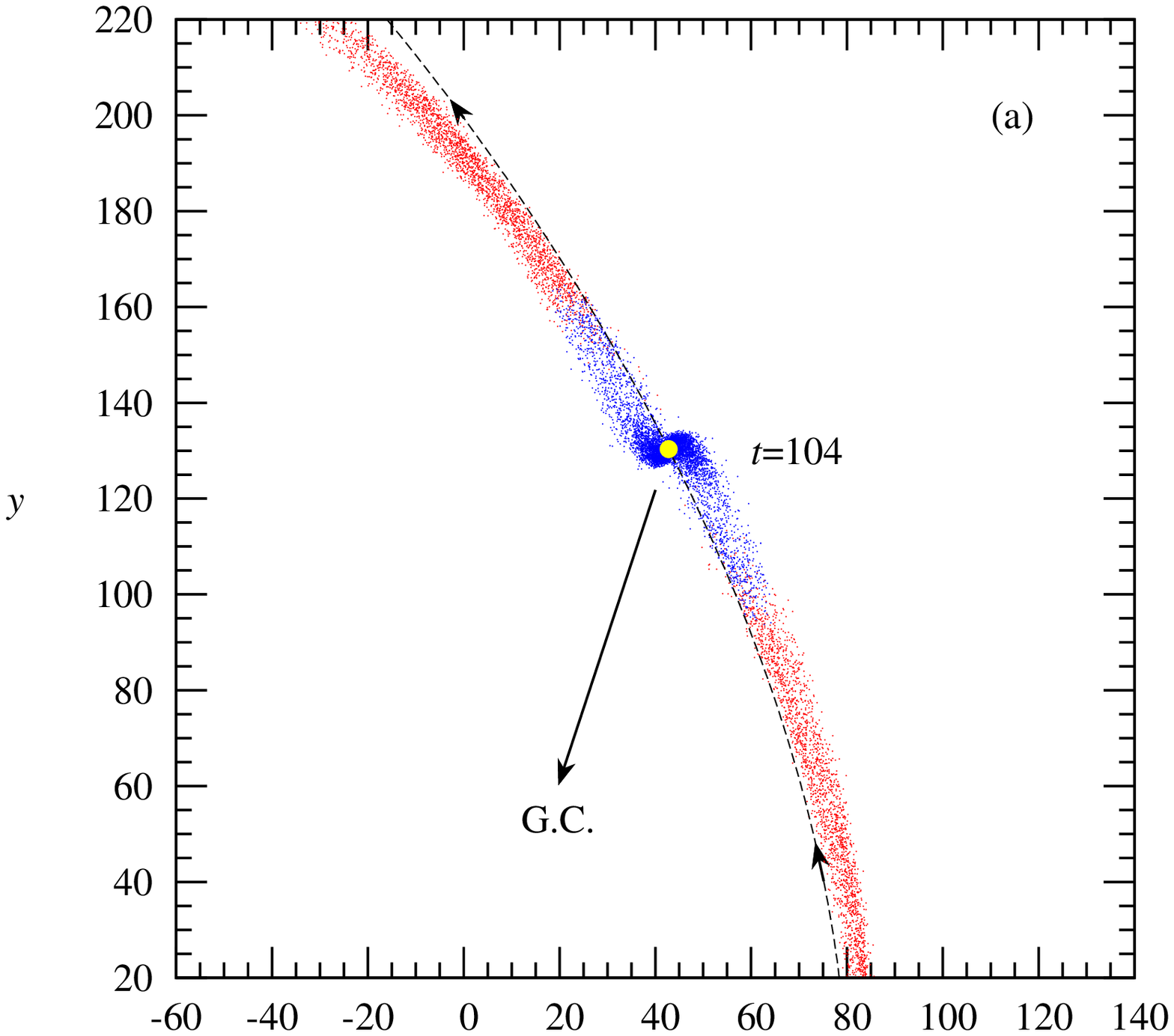}}
\centerline{\includegraphics[width=7cm]{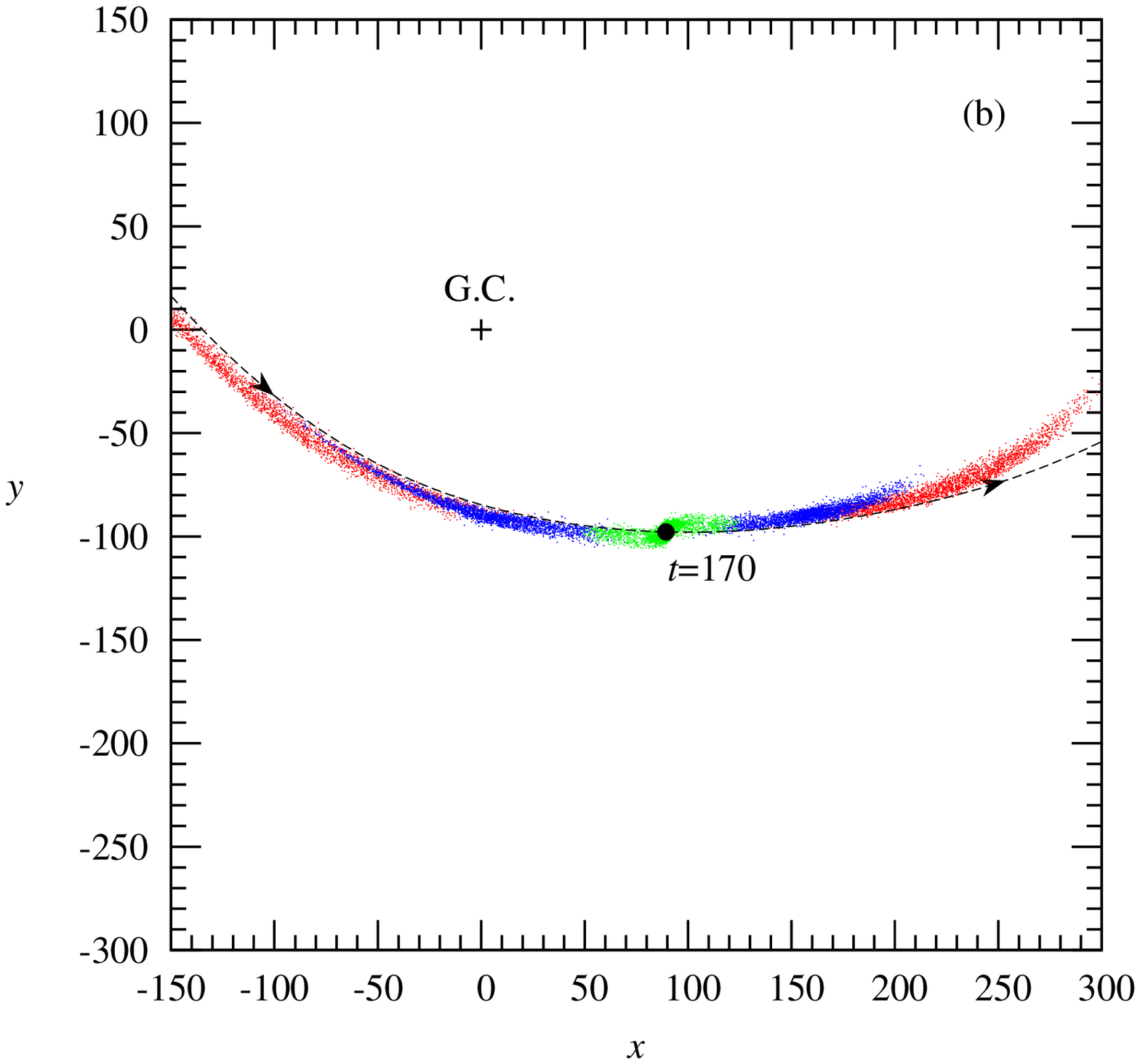}}
\caption{Two pairs of leading and trailing tidal arms
at $t=104$ (a), and three pairs of those at $t=170$ (b),
somewhat after pericentre passage projected on to the
orbital plane. The red, blue, and green dots represent
the particles constituting the firstly, secondly, and
thirdly formed arms, respectively. Only the arm particles
are plotted. The dashed lines show the orbital path
of the dwarf galaxy, while the arrows on them denote
the directions of the orbital motion of the dwarf
galaxy. The filled yellow and black circles indicate
the positions of the centre of the dwarf galaxy at $t=104$,
and at $t=170$, respectively. The long arrow with the label
G.C. in (a) points to the centre of the host galaxy, while
the $+$ symbol in (b) shows the centre of the host galaxy.
The tidal tails are represented on real scales.}
\label{fig:arms}
\end{figure}

\begin{figure}
\centerline{\includegraphics[width=8cm]{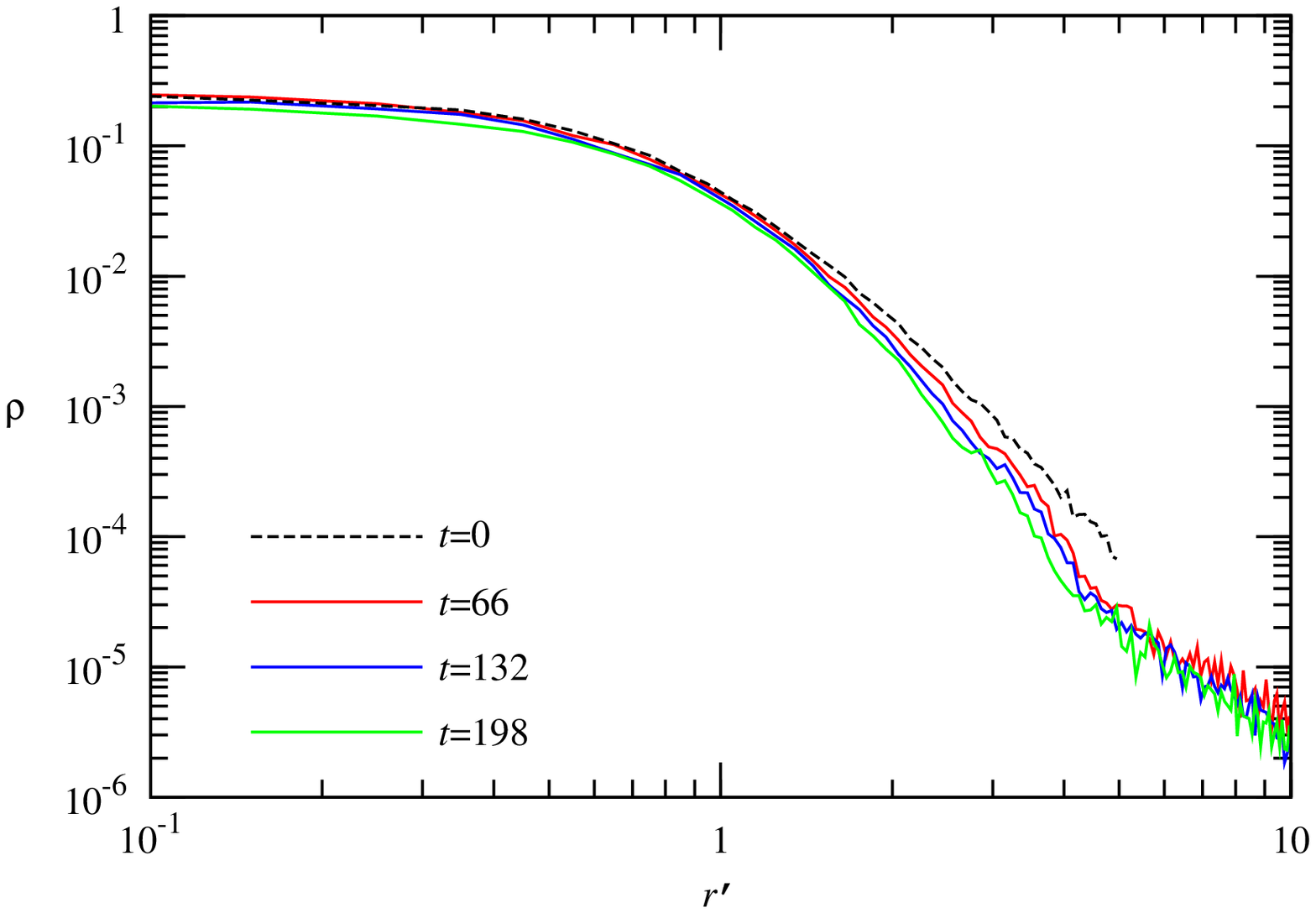}}
\caption{Spherically averaged density distributions of
the main body of the dwarf galaxy around apogalacticons.
Note that the initial model was truncated at $r'=5$.}
\label{fig:body_density}
\end{figure}

Fig.~\ref{fig:armlength} shows the time evolution of the lengths
of leading and trailing arms for the first three consecutively
formed tidal arms. Here, we define the length
of an arm as the linear distance between the nearest and farthest
particles in the arm from the centre of the dwarf galaxy. For
each pair of tidal arms, the lengths of the leading and tailing
arms become longer and longer as the dwarf galaxy moves from an
apogalacticon to the next perigalacticon, and they become shorter
and shorter as it moves from that perigalacticon to the next
apogalacticon. The length of each arm as a whole shows a gradual
increase over time, while this oscillatory behaviour in the arm
length is repeated. The change in the length of
tidal tails has often been mentioned qualitatively
\citep[e.g.][]{lgk13}. This figure demonstrates quantitatively
what we have found from Fig.~\ref{fig:armevolution}: the time
change in the tidal-arm lengths of the secondly formed pair is
one episode behind that of the firstly formed pair, and this
relation between the firstly and secondly formed pairs holds
also between the secondly and thirdly formed pairs.

To illustrate that the three pairs of tidal
tails have similar shapes and density distributions, we first
rotate the positions of the arm particles projected on to the
orbital plane in the secondly formed tails by the angle between
the two apogalacticons at $t\sim 132$ and at $t\sim 198$, and
superpose the secondly formed tails on the firstly formed tails
at the positions of the trailing and leading arms of the firstly
formed tails at $t=132$. Next, these processes are also applied
to the secondly and thirdly formed tails, for which the rotation
angle is the one between the two apogalacticons at $t\sim 198$
and $t\sim 264$, and the superposition is carried out again at
the positions of those arms of the firstly formed tails at
$t=132$. In Fig.~\ref{fig:superpose_arms}, we present such
superpositions of the density contours of the first three
consecutively formed tidal tails, when they are seen in the
orbital plane. From the left panels of
Figs.~\ref{fig:superpose_arms}(a) and \ref{fig:superpose_arms}(b),
we can see that at least, the outlines of the three trailing
and leading arms are in fairly good agreement with one
another. In addition, the density distributions
of the firstly formed tidal tails do not deviate considerably
from those of the secondly formed ones both for the trailing
and leading arms, as illustrated in the middle panels of
Figs.~\ref{fig:superpose_arms}(a) and \ref{fig:superpose_arms}(b).
Likewise, the density distributions of the secondly formed tidal
tails agree to some extent with those of the thirdly formed ones
again both for the trailing and leading arms, as recognized from
the right panels of Figs.~\ref{fig:superpose_arms}(a) and
\ref{fig:superpose_arms}(b).

In Fig.~\ref{fig:arms}, we show the snapshots of the tidal
arms at $t=104$ and at $t=170$, somewhat after pericentre
passage without depicting the main body of the dwarf galaxy.
This figure indicates that at $t=104$, the particles constituting
the first pair of tidal arms are already farther away from the
main body and distributed along either side of its orbital path.
The situation is similar at $t=170$ when both the first and
second arms have disconnected from the main body. As a result,
the region in the vicinity of the main body is always occupied
by the particles that are forming a new pair of tidal arms.
However, as the dwarf galaxy approaches apogalacticon, many
of the constituent particles of the existing pair(s) come
closer to the main body and the lengths of the arms shrink,
as illustrated in Figs.~\ref{fig:tails} and \ref{fig:armlength}.
In this way, the existing pair(s) and the newly formed pair of
tidal arms are found together in the vicinity of the main body
around apocentre passage, as is evident from the snapshots at
$t=132$ and at $t=198$ in Fig.~\ref{fig:tails}. We also notice
that each of the consecutively formed leading and trailing arms
is smoothly connected in line near perigalacticon, which looks
as if only a single pair of tidal arms were created.

As a side note, Fig.~\ref{fig:body_density} presents the
spherically averaged density profiles of the main body
of the dwarf galaxy at the times around apogalacticon.
We have confirmed that the main body itself is roughly
spherical within $r'\sim 10$ until $t\sim 200$.
As seen from this figure, the initial density profile
is deformed outside of $r\sim 2$, the radius of which
is considered to be the tidal radius. Indeed, this radius
can be compared to the tidal radius of $r\sim 1.7$ estimated
at perigalacticon from equation (11) of \citet{ola95}
\citep[see also][]{king62}. The particles that have been
stripped off from the main body produce a noticeable bend
at $r\sim 4$, and continue outward into tidal tails. In
addition, this figure indicates that the tidal forces exerted
by the host galaxy during the formation of the first pair of
tidal arms deform considerably the outer density profile of the
main body, while the subsequent tidal forces have less impact
on the outer density profile, so that it changes gradually
with time. However, the particles included in each pair
of tidal arms is nearly the same in number, regardless
of the strength of the tidal forces, as shown in Table 1.

\section{Discussion}
As far as the formation of the first pair of tidal tails
is concerned, our dwarf galaxy model shows an evolutionary
sequence which is very similar to that in previous studies
\citep{piatek95, dehnen04, capuzzo05, montuori07}, as presented
in Figs.~\ref{fig:evolution} and \ref{fig:armevolution}(a).
That is, the tidal interaction of the dwarf galaxy, starting
from apogalacticon, with its host galaxy generates a pair of
tidal arms at a position close to perigalacticon, and that pair
becomes more pronounced at the subsequent apogalacticon. As
explained qualitatively by \citet{dehnen04} on the basis of
a simple analytic argument on orbital energy, one half of the
pair contains particles which lose energy as a whole and move
on to orbits slightly inside the orbital path of the main body
of the dwarf galaxy, while the other half contains those which
gain energy by and large and move on to orbits slightly outside
that orbital path. Consequently, the former and latter particles
constitute a leading and trailing tail, respectively. In fact,
we have demonstrated in Fig.~\ref{fig:EandJ} that on average,
leading-arm particles lose energy, while trailing-arm
particles gain energy, when a pair of tidal arms is formed.

\begin{table}
\begin{center}
\caption{Number of particles, $N$, and total mass fraction
included in each tidal arm for the first three pairs of tidal
tails. Here, the fraction is the ratio of the number of arm
particles to that of the bound particles of the dwarf galaxy.
The letters L and T denote leading and trailing arms, respectively.}
\begin{tabular}{crrrrrr}
\hline
 & \multicolumn{2}{c}{first arms} & \multicolumn{2}{c}{second arms}
 & \multicolumn{2}{c}{third arms}\\
 & \multicolumn{1}{c}{L} & \multicolumn{1}{c}{T} &
   \multicolumn{1}{c}{L} & \multicolumn{1}{c}{T} &
   \multicolumn{1}{c}{L} & \multicolumn{1}{c}{T}\\
\hline
$N$ & 2$\,$705 & 2$\,$763 & 2$\,$790 & 2$\,$682 & 1$\,$914 & 2$\,$240\\
\%  & 5.41  & 5.53  & 6.27  & 6.02  & 4.90  & 5.73\\
\hline
\end{tabular}
\end{center}
\label{table:armparticles}
\end{table}

On the other hand, we have also shown in Fig.~\ref{fig:EandJ}
that on the whole, leading-arm particles lose angular momentum,
while trailing-arm particles gain angular momentum. After the
arm particles that have lost or gained angular momentum pass
through their first apogalacticon, they lose or gain only a
small amount of angular momentum subsequently. As the dwarf
galaxy moves from an apogalacticon to the next perigalacticon,
the distances from the arm particles to the centre of the host
galaxy diminish, and owing to that approximate conservation of
angular momentum, the velocities of leading-arm particles are
increased while those of trailing-arm particles are decreased
relative to the velocity of the centre of the dwarf galaxy.
Consequently, near the perigalacticon the leading arm
extends forward while the trailing arm elongates backward
with respect to the main body of the dwarf galaxy. Therefore,
as is demonstrated in Figs.~\ref{fig:armevolution} and
\ref{fig:arms}, the direction of the tidal arms becomes
almost parallel to the orbital path of the dwarf galaxy
around pericentre passage, which is often reported in the
literature \citep{dehnen04,capuzzo05,montuori07} without
physical explanations. Subsequently, the arm lengths are
decreased from the perigalacticon to the next apogalacticon
again on account of that approximate conservation of angular
momentum.

Regarding the formation mechanism of multiple tidal arms,
we can see from Figs.~\ref{fig:tails} and \ref{fig:armevolution}
that each pair of tidal arms never bifurcates with time, and
that a fresh pair of tidal arms develops every time the dwarf
galaxy moves from a perigalacticon to the next apogalacticon.
It thus follows that the multiple pairs of tidal arms are
generated by the repeated episode of tidal-arm formation.
However, even though a pair of tidal arms is formed repeatedly
again and again, the multiplicity itself cannot be recognized
unless the angle between two pairs of tidal arms is different
to some degree. Since the dynamical time of the dwarf galaxy
is shorter than its orbital period, it recovers dynamical
equilibrium soon after a pair of tidal arms is extracted
at perigalacticon. Moreover, the gravitational effects of
tidal arms are negligible, so that existing tidal arms do
not affect the subsequent formation of tidal arms. Hence,
the configuration of a forming pair of tidal arms with
respect to the dwarf galaxy is determined only by its
orbital phase. Now that the dwarf galaxy is moving on
an eccentric rosetta orbit that is not
closed, the precession of the orbit results in a rotation
of the orientation of newly forming arms extracted around
perigalacticon, leading to differences in the position
angles of these arm particles. The resulting multiplicity
is recognizable at the apogalacticon where the extracted arms
are well-developed. In this way, the multiplicity of tidal
arms is perceived. In fact, Fig.~\ref{fig:superpose_arms}
proves that the secondly formed pair of tidal arms can be
superposed on the firstly formed pair by rotating the angle
between the two apogalacticons at $t\sim 132$ and $t\sim 198$,
and that likewise, the thirdly formed pair can be fit with the
secondly formed pair by rotating the angle between the two
apogalacticons at $t\sim 198$ and $t\sim 264$. These angles
rotated to superpose the density profiles of tidal tails
correspond to the precessing angle resulting from the eccentric
orbit of the dwarf galaxy. We thus find that multiple tidal
arms are the outcome of repeated tidal-arm formation amid the
precessing motion. Although \citet{montuori07} described the
formation of multiple tidal tails around GCs on the basis of
the Coriolis acceleration term in their equation (5), they
only pointed out the directions of tidal tails near the apo-
and peri-galactic centres, and they never referred to the
precession angle of the orbit.

In our experiment, the number of particles included in
each pair of tidal arms is nearly identical for the first
three pairs of tidal arms as found from Table 1. These
particles produce similar density distributions at similar
orbital positions for the three pairs of tidal arms
(see Fig.~\ref{fig:superpose_arms}). In addition,
Fig.~\ref{fig:EJ_distri} reveals the similarity
in the energy and angular-momentum distributions
of these three pairs. Therefore, each episode of
tidal-arm formation is independent of the previous
episode(s). Again, this fact ensures that the phase
of the tidal-arm formation is specified only by the
phase of the orbit of the dwarf galaxy.

We should discuss the detectability of multiple tidal tails.
First of all, although dwarf galaxies are considered to be
dark-matter dominated \citep{mateo98, gilmore07}, we have
neglected a contribution of dark matter to our dwarf galaxy
model. In such circumstances, it is conceivable that multiple
tidal tails are predominantly made of dark matter. How much
luminous matter is contained in tidal tails can be estimated
from its total fraction outside the tidal radius of a
progenitor dwarf galaxy. Probably, multiple tidal tails
containing stars could also form for plausible models of
dwarf galaxies despite the fact that they are dominated
by dark matter. For example, according to the simulations
of \citet{lgk13} in which the properties of tidal tails
around dwarf galaxies orbiting a Milky Way-like host have
been studied, both stellar and dark-matter components develop
tidal tails. On the other hand, as shown by
\citet{pen08} and by \citet{smith13}, dwarf galaxy models
consisting of stellar and dark-matter components lose more
than 90 per cent of the dark halo before affecting the stellar
component. In this way, even if the stars are not self-gravitating,
they reside in the innermost parts of the dark halo. Consequently,
our one-component model for dwarf galaxies can still provide
some insight into the physics for tidal-tail formation and
the multiplicity in general, and can serve as a basis for
more complex multi-component models. Next, obviously,
we can observe the multiplicity more easily when the orbital
plane of a dwarf galaxy is viewed from the direction to the
line of sight as perpendicularly as possible. However, even
though we are then placed at a preferred configuration with
respect to the orbital plane, Fig.~\ref{fig:arms} indicates
that near perigalacticon, tidal arms generated in subsequent
episodes form a single coherent structure, which makes it
difficult to distinguish each pair of them. As found from
Figs.~\ref{fig:tails} and \ref{fig:armevolution}, the situation
is better around apogalacticon where at least trailing arms are
spatially separated, providing information, in some cases, about
the number of orbital rotations completed by the satellite.
Thus, it is significant to discover those dwarf satellite
galaxies presenting multiple tidal arms which lie around
apogalacticon and whose orbital plane is nearly perpendicular
to the line of sight.

As we have described in Section 1, our present analysis is
applicable to GCs, although we do not include in our study
the effects of disc shocking that would be expected for most
of GCs. Therefore, it is likely that multiple tidal tails
around GCs can be detected. Interestingly, NGC 288 has been
reported to have multiple tidal tails \citep{lmc00}.
However, GCs are considerably less massive than dwarf galaxies,
and so, long tidal tails, in general, would become sufficiently
faint. In addition, since GCs are collisional systems in which
mass segregation progresses, tidal tails consist primarily of
less massive and less luminous stars as shown by \citet{clm99}.
Consequently, it might be more difficult to detect multiple
tidal tails around GCs, even if they actually exist, than
those around dwarf galaxies. As another view, GCs moving
on highly eccentric orbits that are suitable for developing
multiple tidal tails might have been selectively disrupted
on the ground that they populate in the inner regions of
our Galaxy where tidal effects are relatively strong.

\section{Conclusions}
We have examined how tidal tails are formed and evolve to
generate multiple tidal features around a dwarf satellite
galaxy moving on a highly eccentric orbit in a fixed potential
of its host galaxy in which a disc component
is neglected, assuming that the dwarf galaxy is a one-component
system. We find that a pair of tidal tails is excited every
time the dwarf galaxy passes through perigalacticon where it
suffers strong tidal forces from the host galaxy, and that
the extracted pair is added to the existing pair(s) of tidal
tails, which leads to the multiplicity. We also find that the
angle between two consecutively formed pairs of tidal tails
is equal to the one arising from the precessing motion of the
dwarf galaxy, when viewed on the orbital plane. This process
should be valid for any satellite, regardless of whether
it is a GC or a satellite galaxy. Whenever the satellite
is on an eccentric orbit, the repeating tidal-tail formation
process at perigalacticon, coupled with the precessing motion,
should create multiple tidal tails.

In addition, we have demonstrated that especially for the
first three consecutively formed pairs of tidal tails each
pair shows, in effect, identical energy and angular-momentum
distributions as well as density distribution to one another.
This fact indicates that each pair of tidal tails is formed
almost independently of the pre-existing pair(s). It thus turns
out that the phase of the tidal-tail formation is specified
basically by the phase of the eccentric orbit of a satellite
system.

\section*{Acknowledgments}
We are grateful to Drs.\ Andrew Benson, Michael Hilker,
Andreas Koch, and Mike Rich for their critical reading
and valuable comments on the manuscript. We are also grateful
to Drs.\ Masaki Iwasawa and Keigo Nitadori for fruitful
discussions on the angular-momentum changes of arm particles.
We are indebted to the anonymous referee for the valuable
comments that have helped to improve the manuscript. SH
thanks the hospitality of the Max-Planck Institut f\"ur
Astronomie in Heidelberg where this research began and
part of it was completed. AB acknowledges support from
the Cluster of Excellence `Origin and Structure of the
Universe'.

\appendix
\section{Energy and angular-momentum changes for tidal tails at
the formation epoch}

\begin{figure}
\centerline{\includegraphics[width=7.5cm]{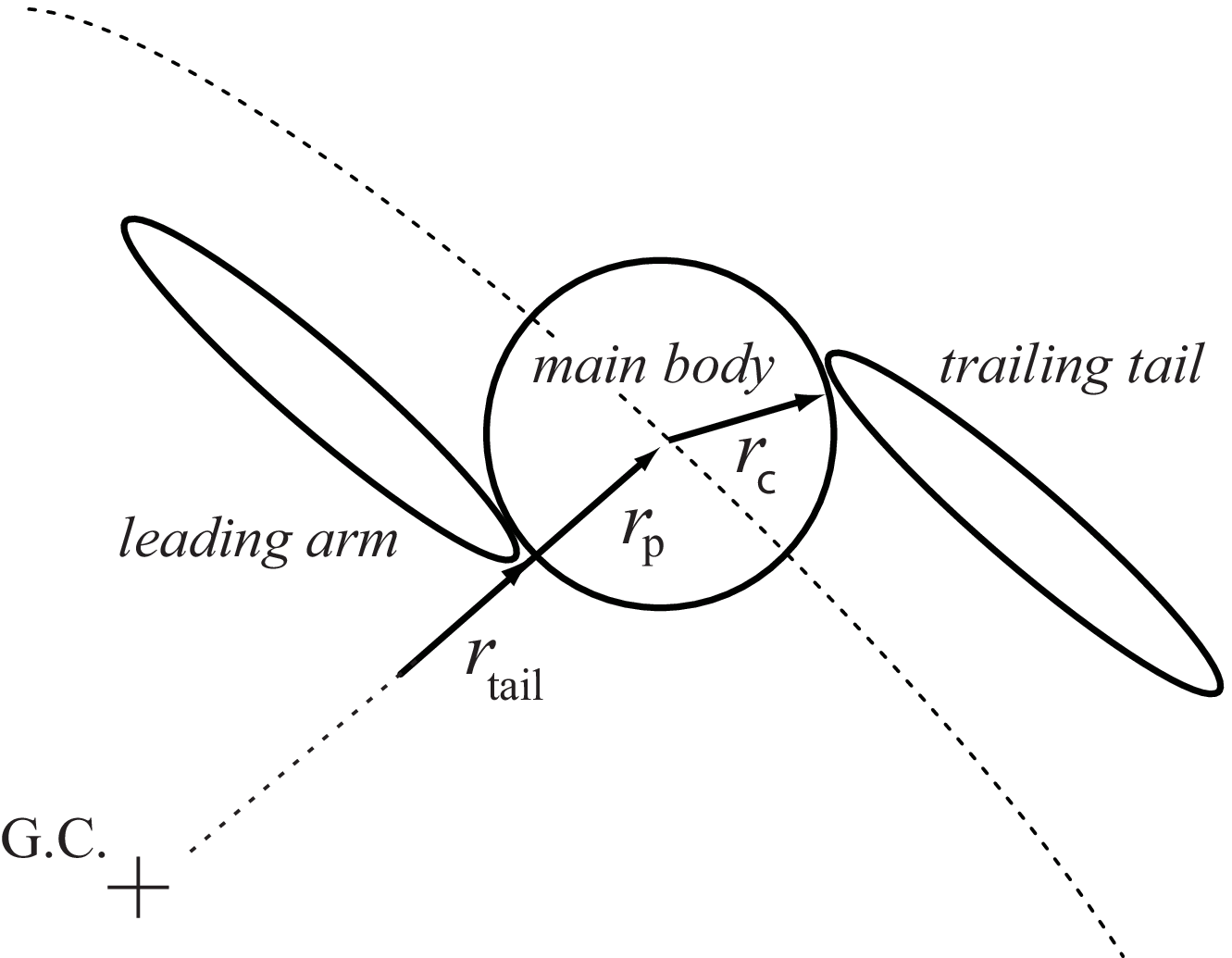}}
\caption{Schematic representation of the main body and the
tidal tails, where $r_{\rm p}$ and $r_{\rm c}$ are the pericentre
distance and the radius of the dwarf galaxy, respectively, and
$r_{\rm tail}$ is a representative distance from the centre of
the host galaxy to the leading or trailing tail.}
\label{fig:energy_change}
\end{figure}

Suppose that the tidal tails are extracted impulsively from
the dwarf galaxy at pericentre. The energy change for
the arm particles arises from the difference in the host
galaxy potential between the distances of the arm particles
from the centre of the host galaxy before and after the
extraction. From Fig.~\ref{fig:energy_change} in which
the configuration of the main body and the tidal tails
is illustrated, the value of this energy change can be
estimated to be $\Delta E = \Phi(r_{\rm tail})-\Phi(r_{{\rm p}})$.
Using equation (\ref{eq:host_pot}), we obtain, in our dimensionless
system of units, $\Delta E = {v_{\rm c}}^2\ln(r_{\rm tail}/r_{\rm p})
= {v_{\rm c}}^2 \ln(1\pm r_{\rm c}/r_{\rm p}) \sim \pm 18$,
where the $+$ and $-$ signs correspond to the trailing and
leading arms, respectively. In this calculation, we adopt
the values of ${v_{\rm c}}^2=282, r_{\rm c}=5$, and $r_{\rm p}=80$.
Thus, $\Delta E$ agrees well with the amount of the energy lost by
the leading arm, or that gained by the trailing arm, as shown in
Fig.~\ref{fig:EandJ}.

\begin{figure}
\centerline{\includegraphics[width=8cm]{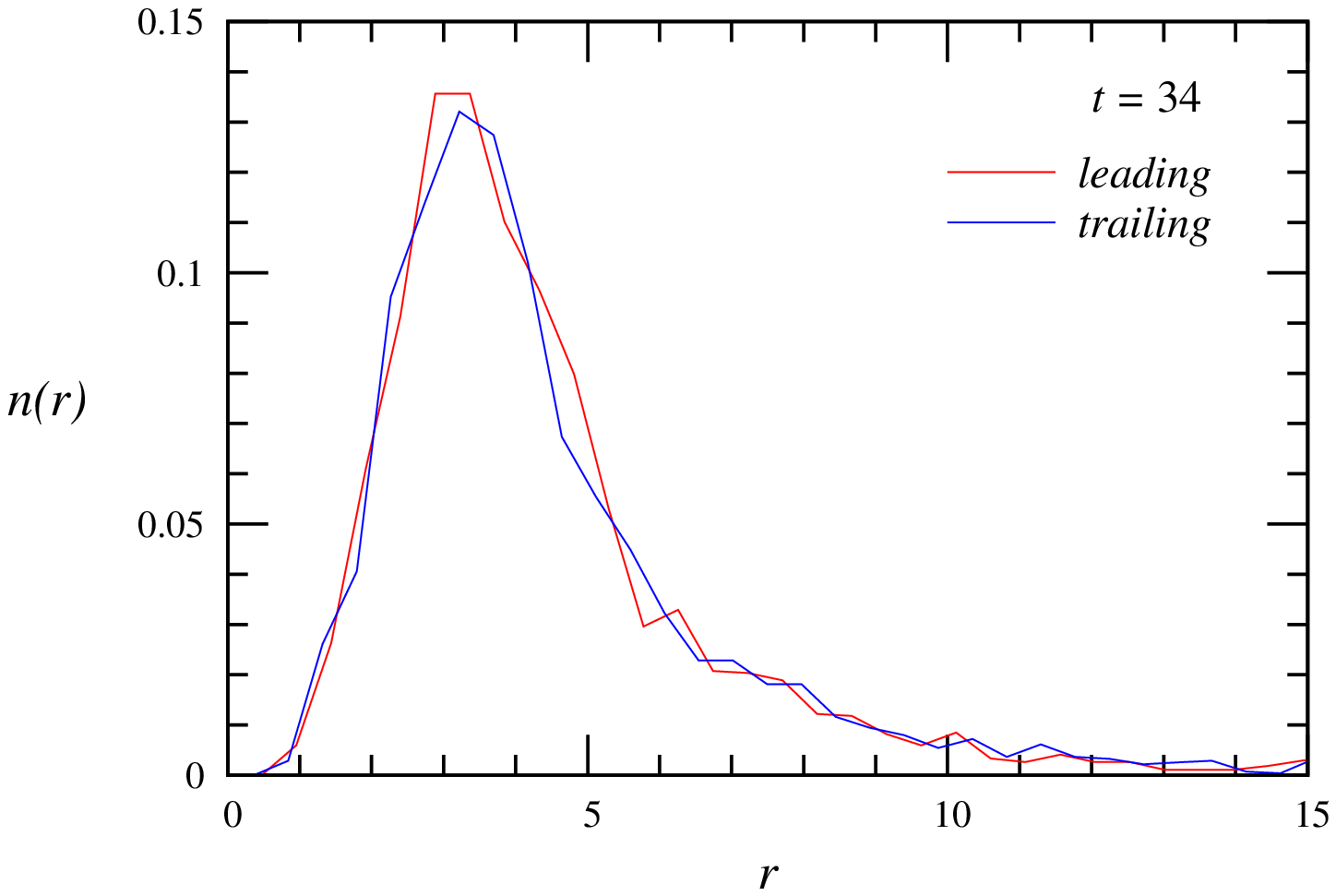}}
\caption{Fractional particle number distributions of the firstly
formed tidal arms against the distance from the centre of the
dwarf galaxy near the pericentre ($t=34$).}
\label{fig:peridistance}
\end{figure}

On the other hand, the angular-momentum change for the arm
particles is caused by stripping them off from the dwarf
galaxy when it passes through the pericentre at the formation
epoch $(t\sim 33)$. Again, this stripping is supposed to
be impulsive, so that the leading-arm particles will lose
angular momentum, while the trailing-arm particles will
gain angular momentum, by the amount of $\Delta L \sim
v_{\rm p}\, \Delta r$, where $v_{\rm p}$ and $\Delta r$
denote, respectively, the orbital speed of the main body
at the pericentre, and the displacement between the centre
of the main body and a representative position of each
arm. In order to evaluate this representative position,
we calculate the distances of the particles in the first
pair of tidal arms from the centre of the dwarf galaxy.
Fig.~\ref{fig:peridistance} shows the resulting fractional
particle number distributions of this pair near the pericentre.
We find from this figure that the typical value of $\Delta r$
is $\sim\pm 3.5$, where the $+$ and $-$ signs correspond
to the trailing and leading arms, respectively. At around
the distance of $|\Delta r|$, the density bend can be found
in Fig.~\ref{fig:body_density}, and so, this distance is
considered to indicate the representative position for both
arms. Consequently, we obtain $\Delta L\sim \pm 105$
using $v_{\rm p}\sim 30$. This value of $\Delta L$ is
close to the amount of the angular momentum lost by the
leading arm, or that gained by the trailing arm, as shown
in Fig.~\ref{fig:EandJ}.

\bsp

\label{lastpage}


\begin{thebibliography}{99}
\bibitem[\protect\citeauthoryear{Balbinot et al.}{2011}]{balbinot11}
  Balbinot E., Santiago B. X., da Costa L. N., Makler M., Maia M. A. G.,
  2011, MNRAS, 416, 393

\bibitem[\protect\citeauthoryear{Barnes \& Hut}{1986}]{bh86}
  Barnes J., Hut P., 1986, Nature, 324, 446

\bibitem[\protect\citeauthoryear{Belokurov et al.}{2006}]{belokurov06}
  Belokurov V., Evans N. W., Irwin M. J., Hewett P. C., Wilkinson M. I.,
  2006, ApJ, 637, L29

\bibitem[\protect\citeauthoryear{Binney}{2008}]{binney08}
  Binney J., 2008, MNRAS, 386, L47

\bibitem[\protect\citeauthoryear{Boylan-Kolchin et al.}{2010}]{bk10}
  Boylan-Kolchin M., Springel V., White S. D. M., Jenkins A., 2010,
  MNRAS, 406, 896

\bibitem[\protect\citeauthoryear{Capuzzo Dolcetta, Di Matteo \& Miocchi}
  {Capuzzo Dolcetta et al.}{2005}]{capuzzo05} Capuzzo Dolcetta R., 
  Di Matteo P., Miocchi P., 2005, AJ, 129, 1906

\bibitem[\protect\citeauthoryear{Choi, Weingerg \& Katz}{2007}]{cwk07}
  Choi J.-H., Weinberg M. D., Katz N., 2007, MNRAS, 381, 987

\bibitem[\protect\citeauthoryear{Coleman et al.}{2005}]{coleman05}
  Coleman M. G., Da Costa G. S., Bland-Hawthorn J., Freeman K. C.,
  2005, AJ, 129, 1443

\bibitem[\protect\citeauthoryear{Combes, Leon \& Meylan}{1999}]{clm99}
  Combes F., Leon S., Meylan G., 1999, A\&A, 352, 149

\bibitem[\protect\citeauthoryear{Dehnen et al.}{2004}]{dehnen04}
  Dehnen W., Odenkirchen M., Grebel E. K., Rix H.-W., 2004, AJ, 127, 2753

\bibitem[\protect\citeauthoryear{Diemand, Kuhlen \& Madau}{2007}]{dkm07}
  Diemand J., Kuhlen M., Madau P., 2007, ApJ, 667, 859

\bibitem[\protect\citeauthoryear{Gilmore et al.}{2007}]{gilmore07}
  Gilmore G., Wilkinson M. I., Wyse R. F. G., Kleyna J. T., Koch A.,
  Evans N. W., Grebel E. K., 2007, ApJ, 663, 948

\bibitem[\protect\citeauthoryear{Grillmair}{1998}]{grillmair98}
  Grillmair C. J., 1998, in Zaritsky D., ed., ASP Conf. Ser. Vol. 136,
  Galactic Halos: A UC Santa Cruz Workshop. Astron. Soc. Pac.,
  San Francisco, p. 45

\bibitem[\protect\citeauthoryear{Grillmair \& Johnson}{2006}]{gj06}
  Grillmair C. J., Johnson R., 2006, ApJ, 639, L17

\bibitem[\protect\citeauthoryear{Grillmair et al.}{Grillmair et al.'s}{1995}]
  {grillmair95} Grillmair C. J., Freeman K. C., Irwin M., Quinn P. J., 1995,
  AJ, 109, 2553

\bibitem[\protect\citeauthoryear{Howley et al.}{2008}]{howley08}
  Howley K. M., Geha M., Guhathakurta P., Montgomery R. M.,
  Laughlin G., Johnston K. V., 2008, ApJ, 683, 722 

\bibitem[\protect\citeauthoryear{King}{1962}]{king62}
  King I., 1962, AJ, 67, 471

\bibitem[\protect\citeauthoryear{Klimentowski et al.}{2009}]{klkmmp09}
  Klimentowski J., {\L}okas E. L., Kazantzidis S., Mayer L., Mamon G. A.,
  Prada F., 2009, MNRAS, 400, 2162

\bibitem[\protect\citeauthoryear{Koch et al.}{2012}]{koch12}
  Koch A., Burkert A., Rich R. M., Collins M. L. M., Black C. S.,
  Hilker M., Benson A. J., 2012, ApJ, 755, L13

\bibitem[\protect\citeauthoryear{K\"upper, Lane \& Heggie}{2012}]{kuep12}
  K\"upper A. H. W., Lane R. R., Heggie D.C., 2012, MNRAS, 420, 2700

\bibitem[\protect\citeauthoryear{Lee et al.}{2003}]{lee03}
  Lee K. H., Lee H. M., Fahlman G. G., Lee M. G., 2003, AJ, 126, 815

\bibitem[\protect\citeauthoryear{Lehmann \& Scholz}{1997}]{lehmann97}
  Lehmann I., Scholz, R.-D., 1997, A\&A, 320, 776

\bibitem[\protect\citeauthoryear{Leon, Meylan \& Combes}{Leon et al.}
  {2000}]{lmc00} Leon S., Meylan G., Combes F., 2000, A\&A, 359, 907

\bibitem[\protect\citeauthoryear{{\L}okas, Gajda \& Kazantzidis}
{{\L}okas et al.}{2013}]{lgk13}
  {\L}okas E. L., Gajda G., Kazantzidis S., 2013, MNRAS, 433, 878

\bibitem[\protect\citeauthoryear{Mart\'inez-Delgado et al.}{2001a}]
  {martinez01a} Mart\'inez-Delgado D., Alonso-Garc\'ia J., Aparicio A.,
   G\'omez-Flechoso M. A., 2001a, ApJ, 549, L63

\bibitem[\protect\citeauthoryear{Mart\'inez-Delgado et al.}{2001b}]
  {martinez01b} Mart\'inez-Delgado D., Aparicio A., G\'omez-Flechoso
   M. \'A., Carrera R., 2001b, ApJ, 549, L199

\bibitem[\protect\citeauthoryear{Mart\'inez-Delgado et al.}{2004}]
  {martinez04} Mart\'inez-Delgado D., G\'omez-Flechoso, M. \'A., Aparicio
  A., Carrera R., 2004, ApJ, 601, 242

\bibitem[\protect\citeauthoryear{Mart\'inez-Delgado et al.}{2012}]
  {martinez12} Mart\'inez-Delgado D. et al., 2012, ApJ, 748, L24

\bibitem[\protect\citeauthoryear{Mateo}{1998}]{mateo98}
  Mateo M., 1998, ARA\&A, 36, 435

\bibitem[\protect\citeauthoryear{Mateo, Olszewski \& Morrison}{1998}]
  {mateoetal98} Mateo M., Olszewski E. W., Morrison H. L., 1998, ApJ,
  508, L55

\bibitem[\protect\citeauthoryear{Montuori et al.}{2007}]
  {montuori07} Montuori M., Capuzzo-Dolcetta R., Di Matteo P.,
  Lepinette A., Miocchi P., 2007, ApJ, 659, 1212

\bibitem[\protect\citeauthoryear{Mu\~noz et al.}{2006}]{munoz06}
  Mu\~noz R. R., et al., 2006, ApJ, 649, 201

\bibitem[\protect\citeauthoryear{Odenkirchen et al.}{2002}]{oden02}
  Odenkirchen M., Grebel E. K., Dehnen W., Rix H.-W., Cudworth K. M.,
  2002, AJ, 124, 1497

\bibitem[\protect\citeauthoryear{Odenkirchen et al.}{2001}]{oden01}
  Odenkirchen M. et al., 2001, ApJ, 548, L165

\bibitem[\protect\citeauthoryear{Odenkirchen et al.}{2003}]{oden03}
  Odenkirchen M. et al., 2003, AJ, 126, 2385

\bibitem[\protect\citeauthoryear{Oh, Lin \& Aarseth}{1995}]{ola95}
  Oh K. S., Lin D. N. C., Aarseth S. J., 1995, ApJ, 442, 142

\bibitem[\protect\citeauthoryear{Pe\~narrubia, Navarro \& McConnachie}{2008}]
  {pen08} Pe\~narrubia J., Navarro J. F., McConnachie A. W., 2008,
  ApJ, 673, 226

\bibitem[\protect\citeauthoryear{Pe\~narrubia et al.}{2009}]{pen09}
  Pe\~narrubia J., Navarro J. F., McConnachie A. W., Martin N. F., 2009,
  ApJ, 698, 222

\bibitem[\protect\citeauthoryear{Piatek \& Pryor}{1995}]{piatek95}
  Piatek S., Pryor C., 1995, AJ, 109, 1071

\bibitem[\protect\citeauthoryear{Plummer}{1911}]{plummer11}
  Plummer H. C., 1911, MNRAS, 71, 460

\bibitem[\protect\citeauthoryear{Press et al.}{1986}]{press86}
  Press W. H., Flannery B. P., Teukolsky S. A., Vetterling, W. T.,
  1986, Numerical Recipes: The Art of Scientific Computing.
  Cambridge Univ. Press, Cambridge, p.~631

\bibitem[\protect\citeauthoryear{Rich et al.}{2012}]{rich12}
  Rich R. M., Collins M. L. M., Black C. M., Longstaff F.A., Koch A.,
  Benson A., Reitzel D. B., 2012, Nature, 482, 192

\bibitem[\protect\citeauthoryear{Saviane, Monaco \& Hallas}{2010}]{sav10}
  Saviane I., Monaco L., Hallas T., 2010, in Bruzual G., Charlot S.,
  eds, Proc. IAU Symp. 262, Stellar Populations -- Planning for the
  Next Decade. Cambridge Univ. Press, Cambridge, p. 426

\bibitem[\protect\citeauthoryear{Siegel et al.}{2001}]{siegel01}
  Siegel M. H., Majewski S. R., Cudworth K. M., Takamiya M.,
  2001, AJ, 121, 935

\bibitem[\protect\citeauthoryear{Smith et al.}{2013}]{smith13}
  Smith R., Fellhauer M., Candlish G. N., Wojtak R., Farias J. P.,
  Bl\~ana M., 2013, MNRAS, 433, 2529

\bibitem[\protect\citeauthoryear{Sofue \& Rubin}{2001}]{sofue01}
  Sofue Y., Rubin V., 2001, ARA\&A, 39, 137

\bibitem[\protect\citeauthoryear{Sohn et al.}{2007}]{sohn07}
  Sohn S. T. et al., 2007, ApJ, 663, 960

\bibitem[\protect\citeauthoryear{Sollima et al.}{2011}]{sollima11}
  Sollima A., Mart\'inez-Delgado D., Valls-Gabaud D., Pe\~narrubia J.,
  2011, ApJ, 726, 47

\bibitem[\protect\citeauthoryear{Testa et al.}{2000}]{testa00}
  Testa V., Zaggia S. R., Andreon S., Longo G., Scaramella R.,
  Djorgovski S. G., de Carvalho R., 2000, A\&A, 356, 127

\end{thebibliography}
\end{document}